\documentclass[longauth]{aa}

\usepackage{graphicx}
\usepackage{natbib}
\usepackage{scalerel}
\usepackage{comment}
\usepackage{ulem}
\usepackage[dvipsnames]{xcolor}

\usepackage{txfonts}

\usepackage[pdfencoding=auto,psdextra]{hyperref}
\hypersetup{
    colorlinks=true,
    linkcolor=blue,
    filecolor=magenta,      
    urlcolor=blue,
    citecolor=blue
}
\newcommand{\hstfull}{\HST\xspace}
\newcommand{\hst}{HST\xspace}

\newcommand{\jwst}{JWST\xspace}
\newcommand{\gaia}{\textit{Gaia}\xspace}
\newcommand{\euclid}{\textit{Euclid}\xspace}
\newcommand{\qfit}{\texttt{QFIT}\xspace}
\newcommand{\radxs}{\texttt{RADXS}\xspace}

\newcommand{\kstwo}{\texttt{KS2}\xspace}

\makeatletter
\renewcommand*\aa@pageof{, page \thepage{} of \pageref*{LastPage}}
\makeatother

\usepackage[utf8]{inputenc}

\usepackage[switch, modulo]{lineno}

\renewcommand{\linenumbers}[0]{}

\usepackage{euclid}

\defcitealias{griggio2026}{G26}

\begin{document}

\title{\Euclid\/: The convective-transition gap of 47\,Tuc\thanks{This paper is published on behalf of the Euclid Consortium}} 

\newcommand{\orcid}[1]{\unskip\protect\href{https://orcid.org/#1}{\protect\includegraphics[width=8pt,clip]{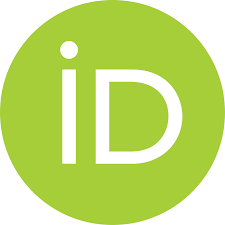}}}
		   
\author{M.~Libralato\orcid{0000-0001-9673-7397}\thanks{\email{mattia.libralato@inaf.it}}\inst{\ref{aff1}}
\and M.~Griggio\orcid{0000-0002-5060-1379}\inst{\ref{aff2}}
\and R.~Gerasimov\orcid{0000-0003-0398-639X}\inst{\ref{aff3}}
\and L.~R.~Bedin\orcid{0000-0003-4080-6466}\inst{\ref{aff1}}
\and I.~McDonald\orcid{0000-0003-0356-0655}\inst{\ref{aff4}}
\and B.~Altieri\orcid{0000-0003-3936-0284}\inst{\ref{aff5}}
\and J.~Anderson\orcid{0000-0003-2861-3995}\inst{\ref{aff2}}
\and F.~Annibali\inst{\ref{aff6}}
\and E.~Balbinot\orcid{0000-0002-1322-3153}\inst{\ref{aff7},\ref{aff8}}
\and G.~Battaglia\orcid{0000-0002-6551-4294}\inst{\ref{aff9},\ref{aff10}}
\and A.~Bellini\orcid{0000-0003-3858-637X}\inst{\ref{aff2}}
\and P.~Casenove\orcid{0009-0006-6736-1670}\inst{\ref{aff11}}
\and J.-C.~Cuillandre\orcid{0000-0002-3263-8645}\inst{\ref{aff12}}
\and E.~Dalessandro\orcid{0000-0003-4237-4601}\inst{\ref{aff6}}
\and A.~M.~N.~Ferguson\inst{\ref{aff13}}
\and H.~R.~A.~Jones\orcid{0000-0003-0433-3665}\inst{\ref{aff14}}
\and K.~Kuijken\orcid{0000-0002-3827-0175}\inst{\ref{aff8}}
\and F.~Z.~Majidi\orcid{0000-0002-8407-5282}\inst{\ref{aff15}}
\and D.~Massari\orcid{0000-0001-8892-4301}\inst{\ref{aff6}}
\and A.~Mohandasan\orcid{0000-0001-5182-0330}\inst{\ref{aff16}}
\and F.~Niederhofer\orcid{0000-0002-4341-9819}\inst{\ref{aff17}}
\and G.~Polenta\orcid{0000-0003-4067-9196}\inst{\ref{aff18}}
\and J.~D.~Sakowska\orcid{0000-0002-1594-1466}\inst{\ref{aff19}}
\and F.~Soldano\inst{\ref{aff20}}
\and C.~Zerbinati\orcid{0009-0004-2797-4056}\inst{\ref{aff21},\ref{aff6}}
\and S.~Andreon\orcid{0000-0002-2041-8784}\inst{\ref{aff22}}
\and N.~Auricchio\orcid{0000-0003-4444-8651}\inst{\ref{aff6}}
\and H.~Aussel\orcid{0000-0002-1371-5705}\inst{\ref{aff12}}
\and C.~Baccigalupi\orcid{0000-0002-8211-1630}\inst{\ref{aff23},\ref{aff24},\ref{aff25},\ref{aff26}}
\and M.~Baldi\orcid{0000-0003-4145-1943}\inst{\ref{aff21},\ref{aff6},\ref{aff27}}
\and A.~Balestra\orcid{0000-0002-6967-261X}\inst{\ref{aff1}}
\and P.~Battaglia\orcid{0000-0002-7337-5909}\inst{\ref{aff6}}
\and A.~Biviano\orcid{0000-0002-0857-0732}\inst{\ref{aff24},\ref{aff23}}
\and E.~Branchini\orcid{0000-0002-0808-6908}\inst{\ref{aff28},\ref{aff29},\ref{aff22}}
\and M.~Brescia\orcid{0000-0001-9506-5680}\inst{\ref{aff30},\ref{aff15}}
\and S.~Camera\orcid{0000-0003-3399-3574}\inst{\ref{aff31},\ref{aff32},\ref{aff33}}
\and V.~Capobianco\orcid{0000-0002-3309-7692}\inst{\ref{aff33}}
\and C.~Carbone\orcid{0000-0003-0125-3563}\inst{\ref{aff34}}
\and J.~Carretero\orcid{0000-0002-3130-0204}\inst{\ref{aff35},\ref{aff36}}
\and M.~Castellano\orcid{0000-0001-9875-8263}\inst{\ref{aff37}}
\and G.~Castignani\orcid{0000-0001-6831-0687}\inst{\ref{aff6}}
\and S.~Cavuoti\orcid{0000-0002-3787-4196}\inst{\ref{aff15},\ref{aff38}}
\and K.~C.~Chambers\orcid{0000-0001-6965-7789}\inst{\ref{aff39}}
\and A.~Cimatti\inst{\ref{aff40}}
\and C.~Colodro-Conde\inst{\ref{aff9}}
\and G.~Congedo\orcid{0000-0003-2508-0046}\inst{\ref{aff13}}
\and C.~J.~Conselice\orcid{0000-0003-1949-7638}\inst{\ref{aff4}}
\and L.~Conversi\orcid{0000-0002-6710-8476}\inst{\ref{aff41},\ref{aff5}}
\and Y.~Copin\orcid{0000-0002-5317-7518}\inst{\ref{aff42}}
\and F.~Courbin\orcid{0000-0003-0758-6510}\inst{\ref{aff43},\ref{aff44},\ref{aff45}}
\and H.~M.~Courtois\orcid{0000-0003-0509-1776}\inst{\ref{aff46}}
\and M.~Cropper\orcid{0000-0003-4571-9468}\inst{\ref{aff47}}
\and H.~Degaudenzi\orcid{0000-0002-5887-6799}\inst{\ref{aff48}}
\and G.~De~Lucia\orcid{0000-0002-6220-9104}\inst{\ref{aff24}}
\and H.~Dole\orcid{0000-0002-9767-3839}\inst{\ref{aff49}}
\and F.~Dubath\orcid{0000-0002-6533-2810}\inst{\ref{aff48}}
\and X.~Dupac\inst{\ref{aff5}}
\and M.~Farina\orcid{0000-0002-3089-7846}\inst{\ref{aff50}}
\and R.~Farinelli\inst{\ref{aff6}}
\and F.~Faustini\orcid{0000-0001-6274-5145}\inst{\ref{aff37}}
\and S.~Ferriol\inst{\ref{aff42}}
\and M.~Frailis\orcid{0000-0002-7400-2135}\inst{\ref{aff24}}
\and E.~Franceschi\orcid{0000-0002-0585-6591}\inst{\ref{aff6}}
\and S.~Galeotta\orcid{0000-0002-3748-5115}\inst{\ref{aff24}}
\and K.~George\orcid{0000-0002-1734-8455}\inst{\ref{aff51}}
\and B.~Gillis\orcid{0000-0002-4478-1270}\inst{\ref{aff13}}
\and C.~Giocoli\orcid{0000-0002-9590-7961}\inst{\ref{aff6},\ref{aff27}}
\and J.~Gracia-Carpio\orcid{0000-0003-4689-3134}\inst{\ref{aff52}}
\and A.~Grazian\orcid{0000-0002-5688-0663}\inst{\ref{aff1}}
\and F.~Grupp\inst{\ref{aff52},\ref{aff53}}
\and S.~V.~H.~Haugan\orcid{0000-0001-9648-7260}\inst{\ref{aff54}}
\and H.~Hoekstra\orcid{0000-0002-0641-3231}\inst{\ref{aff8}}
\and W.~Holmes\orcid{0009-0007-8554-4646}\inst{\ref{aff55}}
\and I.~M.~Hook\orcid{0000-0002-2960-978X}\inst{\ref{aff56}}
\and F.~Hormuth\inst{\ref{aff57}}
\and A.~Hornstrup\orcid{0000-0002-3363-0936}\inst{\ref{aff58},\ref{aff59}}
\and K.~Jahnke\orcid{0000-0003-3804-2137}\inst{\ref{aff60}}
\and M.~Jhabvala\inst{\ref{aff61}}
\and S.~Kermiche\orcid{0000-0002-0302-5735}\inst{\ref{aff62}}
\and A.~Kiessling\orcid{0000-0002-2590-1273}\inst{\ref{aff55}}
\and B.~Kubik\orcid{0009-0006-5823-4880}\inst{\ref{aff42}}
\and M.~K\"ummel\orcid{0000-0003-2791-2117}\inst{\ref{aff53}}
\and M.~Kunz\orcid{0000-0002-3052-7394}\inst{\ref{aff63}}
\and H.~Kurki-Suonio\orcid{0000-0002-4618-3063}\inst{\ref{aff64},\ref{aff65}}
\and A.~M.~C.~Le~Brun\orcid{0000-0002-0936-4594}\inst{\ref{aff66}}
\and S.~Ligori\orcid{0000-0003-4172-4606}\inst{\ref{aff33}}
\and P.~B.~Lilje\orcid{0000-0003-4324-7794}\inst{\ref{aff54}}
\and V.~Lindholm\orcid{0000-0003-2317-5471}\inst{\ref{aff64},\ref{aff65}}
\and I.~Lloro\orcid{0000-0001-5966-1434}\inst{\ref{aff67}}
\and O.~Mansutti\orcid{0000-0001-5758-4658}\inst{\ref{aff24}}
\and O.~Marggraf\orcid{0000-0001-7242-3852}\inst{\ref{aff68}}
\and M.~Martinelli\orcid{0000-0002-6943-7732}\inst{\ref{aff37},\ref{aff69}}
\and N.~Martinet\orcid{0000-0003-2786-7790}\inst{\ref{aff70}}
\and F.~Marulli\orcid{0000-0002-8850-0303}\inst{\ref{aff71},\ref{aff6},\ref{aff27}}
\and R.~J.~Massey\orcid{0000-0002-6085-3780}\inst{\ref{aff72}}
\and E.~Medinaceli\orcid{0000-0002-4040-7783}\inst{\ref{aff6}}
\and S.~Mei\orcid{0000-0002-2849-559X}\inst{\ref{aff73},\ref{aff74}}
\and M.~Meneghetti\orcid{0000-0003-1225-7084}\inst{\ref{aff6},\ref{aff27}}
\and E.~Merlin\orcid{0000-0001-6870-8900}\inst{\ref{aff1}}
\and G.~Meylan\orcid{0000-0001-6503-0209}\inst{\ref{aff75}}
\and A.~Mora\orcid{0000-0002-1922-8529}\inst{\ref{aff76}}
\and L.~Moscardini\orcid{0000-0002-3473-6716}\inst{\ref{aff71},\ref{aff6},\ref{aff27}}
\and R.~Nakajima\orcid{0009-0009-1213-7040}\inst{\ref{aff68}}
\and R.~C.~Nichol\orcid{0000-0003-0939-6518}\inst{\ref{aff77}}
\and S.-M.~Niemi\orcid{0009-0005-0247-0086}\inst{\ref{aff78}}
\and C.~Padilla\orcid{0000-0001-7951-0166}\inst{\ref{aff79}}
\and S.~Paltani\orcid{0000-0002-8108-9179}\inst{\ref{aff48}}
\and F.~Pasian\orcid{0000-0002-4869-3227}\inst{\ref{aff24}}
\and W.~J.~Percival\orcid{0000-0002-0644-5727}\inst{\ref{aff80},\ref{aff81},\ref{aff82}}
\and V.~Pettorino\orcid{0000-0002-4203-9320}\inst{\ref{aff78}}
\and M.~Poncet\inst{\ref{aff11}}
\and L.~A.~Popa\inst{\ref{aff83}}
\and F.~Raison\orcid{0000-0002-7819-6918}\inst{\ref{aff52}}
\and A.~Renzi\orcid{0000-0001-9856-1970}\inst{\ref{aff84},\ref{aff85},\ref{aff6}}
\and J.~Rhodes\orcid{0000-0002-4485-8549}\inst{\ref{aff55}}
\and G.~Riccio\inst{\ref{aff15}}
\and F.~Rizzo\orcid{0000-0002-9407-585X}\inst{\ref{aff24}}
\and E.~Romelli\orcid{0000-0003-3069-9222}\inst{\ref{aff24}}
\and M.~Roncarelli\orcid{0000-0001-9587-7822}\inst{\ref{aff6}}
\and R.~Saglia\orcid{0000-0003-0378-7032}\inst{\ref{aff53},\ref{aff52}}
\and Z.~Sakr\orcid{0000-0002-4823-3757}\inst{\ref{aff86},\ref{aff87},\ref{aff88}}
\and D.~Sapone\orcid{0000-0001-7089-4503}\inst{\ref{aff89}}
\and M.~Schirmer\orcid{0000-0003-2568-9994}\inst{\ref{aff60}}
\and P.~Schneider\orcid{0000-0001-8561-2679}\inst{\ref{aff68}}
\and T.~Schrabback\orcid{0000-0002-6987-7834}\inst{\ref{aff90}}
\and A.~Secroun\orcid{0000-0003-0505-3710}\inst{\ref{aff62}}
\and E.~Sihvola\orcid{0000-0003-1804-7715}\inst{\ref{aff91}}
\and P.~Simon\inst{\ref{aff68}}
\and C.~Sirignano\orcid{0000-0002-0995-7146}\inst{\ref{aff84},\ref{aff85}}
\and G.~Sirri\orcid{0000-0003-2626-2853}\inst{\ref{aff27}}
\and L.~Stanco\orcid{0000-0002-9706-5104}\inst{\ref{aff85}}
\and P.~Tallada-Cresp\'{i}\orcid{0000-0002-1336-8328}\inst{\ref{aff35},\ref{aff36}}
\and A.~N.~Taylor\inst{\ref{aff13}}
\and I.~Tereno\orcid{0000-0002-4537-6218}\inst{\ref{aff92},\ref{aff93}}
\and S.~Toft\orcid{0000-0003-3631-7176}\inst{\ref{aff94},\ref{aff95}}
\and R.~Toledo-Moreo\orcid{0000-0002-2997-4859}\inst{\ref{aff96},\ref{aff97}}
\and F.~Torradeflot\orcid{0000-0003-1160-1517}\inst{\ref{aff36},\ref{aff35}}
\and I.~Tutusaus\orcid{0000-0002-3199-0399}\inst{\ref{aff98},\ref{aff99},\ref{aff87}}
\and J.~Valiviita\orcid{0000-0001-6225-3693}\inst{\ref{aff64},\ref{aff65}}
\and T.~Vassallo\orcid{0000-0001-6512-6358}\inst{\ref{aff24},\ref{aff51}}
\and Y.~Wang\orcid{0000-0002-4749-2984}\inst{\ref{aff100}}
\and J.~Weller\orcid{0000-0002-8282-2010}\inst{\ref{aff53},\ref{aff52}}
\and D.~Scott\orcid{0000-0002-6878-9840}\inst{\ref{aff101}}}
										   
\institute{INAF-Osservatorio Astronomico di Padova, Via dell'Osservatorio 5, 35122 Padova, Italy\label{aff1}
\and
Space Telescope Science Institute, 3700 San Martin Dr, Baltimore, MD 21218, USA\label{aff2}
\and
Department of Physics and Astronomy, University of Notre Dame, Nieuwland Science Hall, Notre Dame, 46556, Indiana, USA\label{aff3}
\and
Jodrell Bank Centre for Astrophysics, Department of Physics and Astronomy, University of Manchester, Oxford Road, Manchester M13 9PL, UK\label{aff4}
\and
ESAC/ESA, Camino Bajo del Castillo, s/n., Urb. Villafranca del Castillo, 28692 Villanueva de la Ca\~nada, Madrid, Spain\label{aff5}
\and
INAF-Osservatorio di Astrofisica e Scienza dello Spazio di Bologna, Via Piero Gobetti 93/3, 40129 Bologna, Italy\label{aff6}
\and
Kapteyn Astronomical Institute, University of Groningen, PO Box 800, 9700 AV Groningen, The Netherlands\label{aff7}
\and
Leiden Observatory, Leiden University, Einsteinweg 55, 2333 CC Leiden, The Netherlands\label{aff8}
\and
Instituto de Astrof\'{\i}sica de Canarias, E-38205 La Laguna, Tenerife, Spain\label{aff9}
\and
Universidad de La Laguna, Dpto. Astrof\'\i sica, E-38206 La Laguna, Tenerife, Spain\label{aff10}
\and
Centre National d'Etudes Spatiales -- Centre spatial de Toulouse, 18 avenue Edouard Belin, 31401 Toulouse Cedex 9, France\label{aff11}
\and
Universit\'e Paris-Saclay, Universit\'e Paris Cit\'e, CEA, CNRS, AIM, 91191, Gif-sur-Yvette, France\label{aff12}
\and
Institute for Astronomy, University of Edinburgh, Royal Observatory, Blackford Hill, Edinburgh EH9 3HJ, UK\label{aff13}
\and
Department of Physics, Astronomy and Mathematics, University of Hertfordshire, College Lane, Hatfield AL10 9AB, UK\label{aff14}
\and
INAF-Osservatorio Astronomico di Capodimonte, Via Moiariello 16, 80131 Napoli, Italy\label{aff15}
\and
Universite Marie et Louis Pasteur, CNRS, Observatoire des Sciences de l'Univers THETA Franche-Comte Bourgogne, Institut UTINAM, Observatoire de Besan\c con, BP 1615, 25010 Besan\c con Cedex, France\label{aff16}
\and
Leibniz-Institut f\"{u}r Astrophysik (AIP), An der Sternwarte 16, 14482 Potsdam, Germany\label{aff17}
\and
Space Science Data Center, Italian Space Agency, via del Politecnico snc, 00133 Roma, Italy\label{aff18}
\and
Instituto de Astrof\'isica de Andaluc\'ia, CSIC, Glorieta de la Astronom\'\i a, 18080, Granada, Spain\label{aff19}
\and
Institut d'Astrophysique de Paris, UMR 7095, CNRS, and Sorbonne Universit\'e, 98 bis boulevard Arago, 75014 Paris, France\label{aff20}
\and
Dipartimento di Fisica e Astronomia, Universit\`a di Bologna, Via Gobetti 93/2, 40129 Bologna, Italy\label{aff21}
\and
INAF-Osservatorio Astronomico di Brera, Via Brera 28, 20122 Milano, Italy\label{aff22}
\and
IFPU, Institute for Fundamental Physics of the Universe, via Beirut 2, 34151 Trieste, Italy\label{aff23}
\and
INAF-Osservatorio Astronomico di Trieste, Via G. B. Tiepolo 11, 34143 Trieste, Italy\label{aff24}
\and
INFN, Sezione di Trieste, Via Valerio 2, 34127 Trieste TS, Italy\label{aff25}
\and
SISSA, International School for Advanced Studies, Via Bonomea 265, 34136 Trieste TS, Italy\label{aff26}
\and
INFN-Sezione di Bologna, Viale Berti Pichat 6/2, 40127 Bologna, Italy\label{aff27}
\and
Dipartimento di Fisica, Universit\`a di Genova, Via Dodecaneso 33, 16146, Genova, Italy\label{aff28}
\and
INFN-Sezione di Genova, Via Dodecaneso 33, 16146, Genova, Italy\label{aff29}
\and
Department of Physics "E. Pancini", University Federico II, Via Cinthia 6, 80126, Napoli, Italy\label{aff30}
\and
Dipartimento di Fisica, Universit\`a degli Studi di Torino, Via P. Giuria 1, 10125 Torino, Italy\label{aff31}
\and
INFN-Sezione di Torino, Via P. Giuria 1, 10125 Torino, Italy\label{aff32}
\and
INAF-Osservatorio Astrofisico di Torino, Via Osservatorio 20, 10025 Pino Torinese (TO), Italy\label{aff33}
\and
INAF-IASF Milano, Via Alfonso Corti 12, 20133 Milano, Italy\label{aff34}
\and
Centro de Investigaciones Energ\'eticas, Medioambientales y Tecnol\'ogicas (CIEMAT), Avenida Complutense 40, 28040 Madrid, Spain\label{aff35}
\and
Port d'Informaci\'{o} Cient\'{i}fica, Campus UAB, C. Albareda s/n, 08193 Bellaterra (Barcelona), Spain\label{aff36}
\and
INAF-Osservatorio Astronomico di Roma, Via Frascati 33, 00078 Monteporzio Catone, Italy\label{aff37}
\and
INFN section of Naples, Via Cinthia 6, 80126, Napoli, Italy\label{aff38}
\and
Institute for Astronomy, University of Hawaii, 2680 Woodlawn Drive, Honolulu, HI 96822, USA\label{aff39}
\and
Dipartimento di Fisica e Astronomia "Augusto Righi" - Alma Mater Studiorum Universit\`a di Bologna, Viale Berti Pichat 6/2, 40127 Bologna, Italy\label{aff40}
\and
European Space Agency/ESRIN, Largo Galileo Galilei 1, 00044 Frascati, Roma, Italy\label{aff41}
\and
Universit\'e Claude Bernard Lyon 1, CNRS/IN2P3, IP2I Lyon, UMR 5822, Villeurbanne, F-69100, France\label{aff42}
\and
Institut de Ci\`{e}ncies del Cosmos (ICCUB), Universitat de Barcelona (IEEC-UB), Mart\'{i} i Franqu\`{e}s 1, 08028 Barcelona, Spain\label{aff43}
\and
Instituci\'o Catalana de Recerca i Estudis Avan\c{c}ats (ICREA), Passeig de Llu\'{\i}s Companys 23, 08010 Barcelona, Spain\label{aff44}
\and
Institut de Ciencies de l'Espai (IEEC-CSIC), Campus UAB, Carrer de Can Magrans, s/n Cerdanyola del Vall\'es, 08193 Barcelona, Spain\label{aff45}
\and
UCB Lyon 1, CNRS/IN2P3, IUF, IP2I Lyon, 4 rue Enrico Fermi, 69622 Villeurbanne, France\label{aff46}
\and
Mullard Space Science Laboratory, University College London, Holmbury St Mary, Dorking, Surrey RH5 6NT, UK\label{aff47}
\and
Department of Astronomy, University of Geneva, ch. d'Ecogia 16, 1290 Versoix, Switzerland\label{aff48}
\and
Universit\'e Paris-Saclay, CNRS, Institut d'astrophysique spatiale, 91405, Orsay, France\label{aff49}
\and
INAF-Istituto di Astrofisica e Planetologia Spaziali, via del Fosso del Cavaliere, 100, 00100 Roma, Italy\label{aff50}
\and
University Observatory, LMU Faculty of Physics, Scheinerstr.~1, 81679 Munich, Germany\label{aff51}
\and
Max Planck Institute for Extraterrestrial Physics, Giessenbachstr. 1, 85748 Garching, Germany\label{aff52}
\and
Universit\"ats-Sternwarte M\"unchen, Fakult\"at f\"ur Physik, Ludwig-Maximilians-Universit\"at M\"unchen, Scheinerstr.~1, 81679 M\"unchen, Germany\label{aff53}
\and
Institute of Theoretical Astrophysics, University of Oslo, P.O. Box 1029 Blindern, 0315 Oslo, Norway\label{aff54}
\and
Jet Propulsion Laboratory, California Institute of Technology, 4800 Oak Grove Drive, Pasadena, CA, 91109, USA\label{aff55}
\and
Department of Physics, Lancaster University, Lancaster, LA1 4YB, UK\label{aff56}
\and
Felix Hormuth Engineering, Goethestr. 17, 69181 Leimen, Germany\label{aff57}
\and
Technical University of Denmark, Elektrovej 327, 2800 Kgs. Lyngby, Denmark\label{aff58}
\and
Cosmic Dawn Center (DAWN), Denmark\label{aff59}
\and
Max-Planck-Institut f\"ur Astronomie, K\"onigstuhl 17, 69117 Heidelberg, Germany\label{aff60}
\and
NASA Goddard Space Flight Center, Greenbelt, MD 20771, USA\label{aff61}
\and
Aix-Marseille Universit\'e, CNRS/IN2P3, CPPM, Marseille, France\label{aff62}
\and
Universit\'e de Gen\`eve, D\'epartement de Physique Th\'eorique and Centre for Astroparticle Physics, 24 quai Ernest-Ansermet, CH-1211 Gen\`eve 4, Switzerland\label{aff63}
\and
Department of Physics, P.O. Box 64, University of Helsinki, 00014 Helsinki, Finland\label{aff64}
\and
Helsinki Institute of Physics, Gustaf H{\"a}llstr{\"o}min katu 2, University of Helsinki, 00014 Helsinki, Finland\label{aff65}
\and
Laboratoire d'etude de l'Univers et des phenomenes eXtremes, Observatoire de Paris, Universit\'e PSL, Sorbonne Universit\'e, CNRS, 92190 Meudon, France\label{aff66}
\and
SKAO, Jodrell Bank, Lower Withington, Macclesfield SK11 9FT, UK\label{aff67}
\and
Universit\"at Bonn, Argelander-Institut f\"ur Astronomie, Auf dem H\"ugel 71, 53121 Bonn, Germany\label{aff68}
\and
INFN-Sezione di Roma, Piazzale Aldo Moro, 2 - c/o Dipartimento di Fisica, Edificio G. Marconi, 00185 Roma, Italy\label{aff69}
\and
Aix-Marseille Universit\'e, CNRS, CNES, LAM, Marseille, France\label{aff70}
\and
Dipartimento di Fisica e Astronomia "Augusto Righi" - Alma Mater Studiorum Universit\`a di Bologna, via Piero Gobetti 93/2, 40129 Bologna, Italy\label{aff71}
\and
Department of Physics, Institute for Computational Cosmology, Durham University, South Road, Durham, DH1 3LE, UK\label{aff72}
\and
Universit\'e Paris Cit\'e, CNRS, Astroparticule et Cosmologie, 75013 Paris, France\label{aff73}
\and
CNRS-UCB International Research Laboratory, Centre Pierre Bin\'etruy, IRL2007, CPB-IN2P3, Berkeley, USA\label{aff74}
\and
Institute of Physics, Laboratory of Astrophysics, Ecole Polytechnique F\'ed\'erale de Lausanne (EPFL), Observatoire de Sauverny, 1290 Versoix, Switzerland\label{aff75}
\and
Telespazio UK S.L. for European Space Agency (ESA), Camino bajo del Castillo, s/n, Urbanizacion Villafranca del Castillo, Villanueva de la Ca\~nada, 28692 Madrid, Spain\label{aff76}
\and
School of Mathematics and Physics, University of Surrey, Guildford, Surrey, GU2 7XH, UK\label{aff77}
\and
European Space Agency/ESTEC, Keplerlaan 1, 2201 AZ Noordwijk, The Netherlands\label{aff78}
\and
Institut de F\'{i}sica d'Altes Energies (IFAE), The Barcelona Institute of Science and Technology, Campus UAB, 08193 Bellaterra (Barcelona), Spain\label{aff79}
\and
Waterloo Centre for Astrophysics, University of Waterloo, Waterloo, Ontario N2L 3G1, Canada\label{aff80}
\and
Department of Physics and Astronomy, University of Waterloo, Waterloo, Ontario N2L 3G1, Canada\label{aff81}
\and
Perimeter Institute for Theoretical Physics, Waterloo, Ontario N2L 2Y5, Canada\label{aff82}
\and
Institute of Space Science, Str. Atomistilor, nr. 409 M\u{a}gurele, Ilfov, 077125, Romania\label{aff83}
\and
Dipartimento di Fisica e Astronomia "G. Galilei", Universit\`a di Padova, Via Marzolo 8, 35131 Padova, Italy\label{aff84}
\and
INFN-Padova, Via Marzolo 8, 35131 Padova, Italy\label{aff85}
\and
Instituto de F\'isica Te\'orica UAM-CSIC, Campus de Cantoblanco, 28049 Madrid, Spain\label{aff86}
\and
Institut de Recherche en Astrophysique et Plan\'etologie (IRAP), Universit\'e de Toulouse, CNRS, UPS, CNES, 14 Av. Edouard Belin, 31400 Toulouse, France\label{aff87}
\and
Universit\'e St Joseph; Faculty of Sciences, Beirut, Lebanon\label{aff88}
\and
Departamento de F\'isica, FCFM, Universidad de Chile, Blanco Encalada 2008, Santiago, Chile\label{aff89}
\and
Universit\"at Innsbruck, Institut f\"ur Astro- und Teilchenphysik, Technikerstr. 25/8, 6020 Innsbruck, Austria\label{aff90}
\and
Department of Physics and Helsinki Institute of Physics, Gustaf H\"allstr\"omin katu 2, University of Helsinki, 00014 Helsinki, Finland\label{aff91}
\and
Departamento de F\'isica, Faculdade de Ci\^encias, Universidade de Lisboa, Edif\'icio C8, Campo Grande, PT1749-016 Lisboa, Portugal\label{aff92}
\and
Instituto de Astrof\'isica e Ci\^encias do Espa\c{c}o, Faculdade de Ci\^encias, Universidade de Lisboa, Tapada da Ajuda, 1349-018 Lisboa, Portugal\label{aff93}
\and
Cosmic Dawn Center (DAWN)\label{aff94}
\and
Niels Bohr Institute, University of Copenhagen, Jagtvej 128, 2200 Copenhagen, Denmark\label{aff95}
\and
Universidad Polit\'ecnica de Cartagena, Departamento de Electr\'onica y Tecnolog\'ia de Computadoras,  Plaza del Hospital 1, 30202 Cartagena, Spain\label{aff96}
\and
European University of Technology EUt+, European Union\label{aff97}
\and
Institute of Space Sciences (ICE, CSIC), Campus UAB, Carrer de Can Magrans, s/n, 08193 Barcelona, Spain\label{aff98}
\and
Institut d'Estudis Espacials de Catalunya (IEEC),  Edifici RDIT, Campus UPC, 08860 Castelldefels, Barcelona, Spain\label{aff99}
\and
Caltech/IPAC, 1200 E. California Blvd., Pasadena, CA 91125, USA\label{aff100}
\and
Department of Physics and Astronomy, University of British Columbia, Vancouver, BC V6T 1Z1, Canada\label{aff101}}    

\date{Received 02 June 2026 / Accepted 25 June 2026}

\abstract{We report the first detection of the `convective-transition gap' (also known as `M-dwarf gap') in the globular cluster 47\,Tuc (NGC\,104) thanks to \euclid data. This feature, linked to a change in the physical properties of late-type dwarfs, has remained elusive, with only two detections so far. Leveraging the large number of stars, high resolution, and photometric precision enabled by \euclid, we detect a statistically significant, sharp discontinuity in the main-sequence luminosity function of 47\,Tuc at \IE$\approx$\,22.9, which we identify as the convective-transition gap. We compare the observed properties of the gap in 47\,Tuc with theoretical models, showing how the gap can be a powerful diagnostic to probe the internal chemical structure of globular clusters, and their multiple stellar populations. Following its initial discovery in the metal-poor cluster NGC\,6397, the identification of a convective gap in the metal-rich 47\,Tuc suggests that this feature might be more general than previously thought. These results demonstrate that \euclid can be transformative well beyond cosmology, with impact across multiple areas of astrophysics, including resolved stellar populations.\looseness=-4}


\titlerunning{\Euclid\/: The convective-transition gap of 47\,Tuc}
\authorrunning{M. Libralato et al.}
   
\maketitle

\section{Introduction}\label{sec:intro}

Although globular clusters (GCs) have been studied for centuries, new telescopes can still yield surprising discoveries. One of the most recent and exciting findings is the detection of the so-called `convective-transition gap' (hereafter `convective gap' or simply `gap'; also known as the `\gaia gap', `M-dwarf gap' or `Jao gap') in a GC \citep[hereafter G26]{griggio2026}. This gap is an elusive feature present in colour-magnitude diagrams (CMDs) of resolved stellar populations, which is attributed to a change in the internal properties (the transition from partial to full convection, and $^3$He production and mixing) of late-type dwarfs \citep{convective_kissing,2018A&A...619A.177B}.\looseness=-4

The convective gap was first identified by \citet{2018ApJ...861L..11J} using the \gaia Data Release 2 \citep{2016GaiaCit,2018A&A...616A...1G} catalogue \citep[see also][]{2021ApJ...907...53F}. Their analysis of solar-neighbourhood dwarfs revealed a narrow, distinct drop in stellar counts along the lower main sequence (MS) at about 0.35\,$M_\odot$. More recently, \citet{2026arXiv260221882M} found a discontinuity in the MS of the open cluster NGC~2158, which they tentatively associated with the same mechanism responsible for the gap. They also suggested that a lack of comparable detections in GCs -- despite numerous \hstfull (\hst) and \jwst programmes -- implied that the gap might be confined to young- or intermediate-age populations. This hypothesis contrasts with the predictions of \citet{2021A&A...650A.184M}, who suggested that GCs are instead ideal systems for studying the gap, and with the recent detection of the gap in the metal-poor ($[\mathrm{Fe/H}] = -1.9$; \citealt{2018ApJ...864..147C}) GC NGC\,6397 (\citetalias{griggio2026}), enabled by the high-precision, wide-field \euclid \citep{EuclidSkyOverview} data of the Early Release Observations \citep[ERO;][]{EROData}.\looseness=-4

Detecting the gap in stellar clusters constrains theoretical models: since asteroseismology is challenging for low-mass stars \citep[e.g.,][]{2019FrASS...6...76R}, the gap morphology may be the most direct observable of the fundamental physical processes that govern the interiors and evolution of these objects. Unlike the heterogeneous \gaia sample, clusters -- being composed of nearly-coeval stars with (almost) the same metallicity -- are ideal benchmarks that simplify the interpretation and theoretical calibration of this feature. However, with only one statistically robust finding to date, characterisation of the gap's underlying physics remains limited.\looseness=-4

This paper reports a clear detection by \euclid of the convective gap in the GC NGC\,104 (hereafter 47\,Tuc), a significantly more metal-rich GC ($[\mathrm{Fe/H}] = -0.75$; \citealt{roman_47tuc}) than NGC\,6397. This suggests the feature might be universal, having been previously missed only due to data limitations.\looseness=-4

\begin{figure*}
    \centering
    \includegraphics[width=\textwidth]{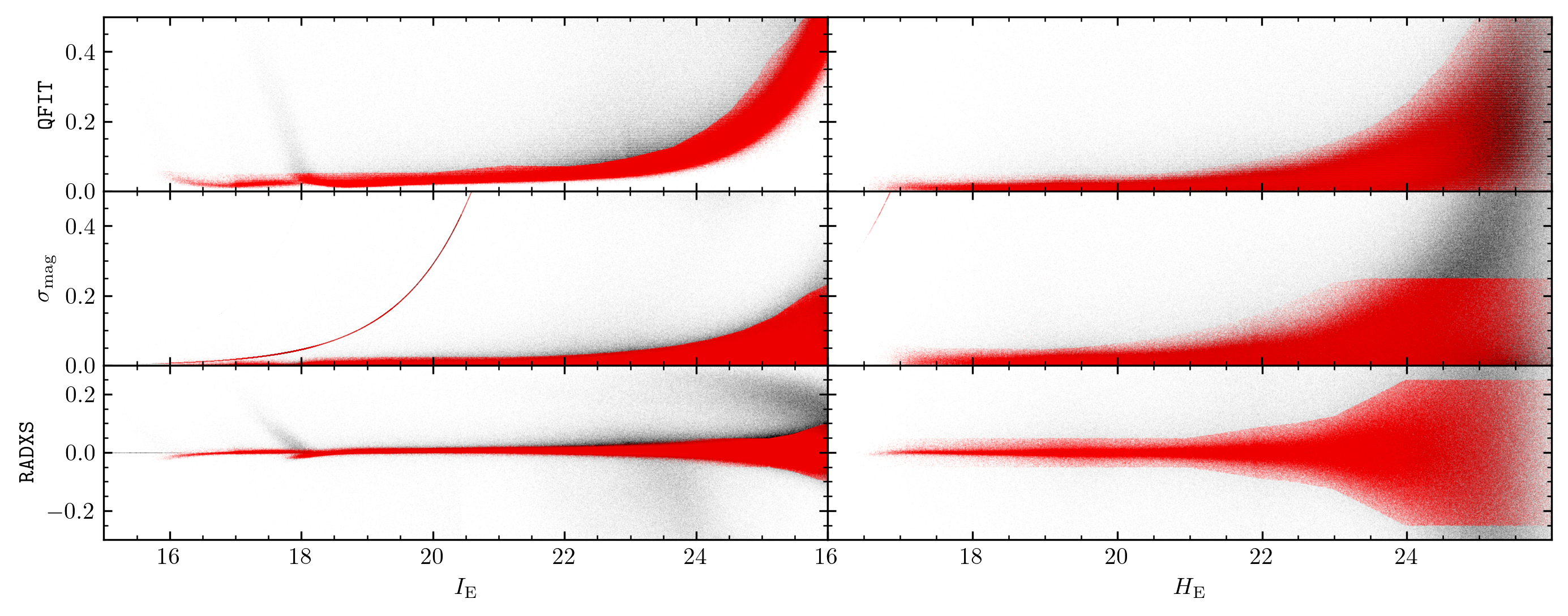}
    \caption{Photometric selections applied to the \IE (left column) and \HE (right column) data. From top to bottom, we plot the \texttt{QFIT} parameter, magnitude root-mean square uncertainty $\sigma_{\rm mag}$, and \texttt{RADXS} parameter as a function of calibrated magnitude. Stars that passed all our photometric criteria are shown in red, while all other sources are plot in black. In the VIS panels, the change in the average trends at $\IE<18$ marks the transition from long to short exposures. The sequence of outliers at $\IE \approx 18$ is likely due to a combination of `brighter-fatter', residual uncorrected non-linearity effects, or some saturated pixels not properly flagged as such in our reduction, which change the shape of the PSF for these sources \citep[see also][]{Libralato24}. The narrow sequences visible in the \IE and \HE magnitude $\sigma_{\rm mag}$ panels include stars measured in only one exposure (i.e., no $\sigma_{\rm mag}$ measurement is possible) for which our photometric tool \kstwo provides a fake exponential trend related to the signal-to-noise ratio of the sources.\looseness=-4}
    \label{fig:selections}
\end{figure*}

\section{Observations and data reduction}\label{sec:datared}

The field including 47\,Tuc represents an example of a part of the Euclid Wide Survey that is of low quality for cosmology, but of high legacy value for stellar populations \citep{Scaramella-EP1,EROGalGCs}. These \euclid observations were secured on 7--8 November 2024 in a 2$\times$2 mosaic centred at about (RA, Dec) $=$ (6.38245, $-$71.86985) deg. Each pointing comprises six (four 560\,s and two 89.5\,s) VIS  \citep{EuclidSkyVIS} exposures (\IE filter) and four (87.2\,s) NISP \citep{EuclidSkyNISP} exposures for each of its three filters (\YE, \JE, \HE). We used Level-2 individual calibrated frames (pipeline-processed, not resampled images), which are best suited for high-precision astrometry and photometry using point-spread-function (PSF) fitting techniques \citep[e.g.,][]{Libralato24}. The data were processed using the official pipeline \citep{Q1-TP002,Q1-TP003} available at the time of writing \citep[`on-the-fly' data;][]{EuclidSkyOverview}.\looseness=-4

The photometry of 47\,Tuc was obtained using the multiple-pass tool described in \citetalias{griggio2026}, which is based on state-of-the-art techniques developed for \hst \citep[e.g.,][]{2008AJ....135.2055A,2017BelliniwCenI} and recently adapted to \euclid (\citealt{Libralato24}; \citetalias{griggio2026}). We initially created a set of positions and fluxes for the brightest and most isolated sources in each \euclid image via effective-PSF (ePSF) fitting. The ePSFs were tailored for each image to account for temporal variations of the ePSF starting from the ePSF models made by \citet{Libralato24}. Positions were corrected with the geometric-distortion correction described in \citet{Libralato24}. Linear terms of the distortion were updated using the \gaia Data Release 3 (DR3) catalogue \citep{2023A&A...674A...1G}, and positions of all images were projected (always using the \gaia DR3 catalogue as described in \citetalias{griggio2026}) onto a common tangent plane with tangent point at the approximate centre of the \euclid mosaic. Bright, unsaturated stars in each catalogue were cross-matched against those in the \gaia DR3 catalogue to set up a common reference-frame system with a defined, yet arbitrary, axis orientation ($X$ and $Y$ axes pointing toward west and north, respectively), and pixel scale (100\,mas\,pixel$^{-1}$). We used these ePSF models, geometric-distortion corrections and reference-frame system to run a \euclid version of the multiple-pass photometry tool \kstwo presented in \citetalias{griggio2026}.\looseness=-4

\kstwo brings two specific features to the analysis: it uses all images at once to improve the detection of faint sources (which would be lost in the noise of a single image); and it performs neighbour subtraction prior to measuring the position and flux of each given object. The latter step is of critical importance in crowded environments like GCs. Photometry is obtained by \kstwo in two ways: ePSF fitting; and fixed-position aperture photometry weighted for the ePSF model. We refer to \citetalias{griggio2026} for more details \citep[but see also][for a description of the implementation of the photometric methods]{2017BelliniwCenI}. Empirical tests have shown that the former method is best with VIS data, while the latter method is preferred for NISP images. Instrumental magnitudes were calibrated in the AB system by computing the magnitude zero-point with the official Level-2 stack calibrated photometry produced by the official \euclid pipeline \citep{EuclidSkyOverview}.

\kstwo also provides quality parameters, which can be used to select well-measured stars. In our work, we considered as well-measured all sources passing the following criteria in both \IE and \HE data.
\begin{itemize} 
    \item The star is not saturated.
    \item The quality of the PSF fit (\qfit) value is lower than the 97.25-th percentile of the distribution at any given magnitude. The closer the \qfit is to zero, the better the ePSF fit. We kept all stars with \qfit\,$<$\,0.05 and rejected all measurements with \qfit\,$>$\,0.5.
    \item The magnitude root-mean square uncertainty $\sigma_{\rm mag}$ is lower than the 97.25-th percentile of the distribution at any given magnitude. We retained all sources with a magnitude $\sigma_{\rm mag}<0.025$\,mag, and discarded those with $\sigma_{\rm mag} > 0.25$\,mag.
    \item The absolute value of the normalised excess/deficiency of flux outside the core of the star with respect to what predicted by the ePSF \citep[hereafter \radxs;][]{2008ApJ...678.1279B} is lower than the 90-th percentile of the distribution at any given magnitude. An object with a $|$\texttt{RADXS}$| < 0.05$ is always included in the well-measured sample, while all sources with $|$\texttt{RADXS}$| > 0.25$ are excluded.
    \item The rejection rate (ratio between the number of images used to measure the flux of a source over the total number of images where the source was found) is lower than 40\%.
    \item The fractional flux within the fitting radius prior to neighbour subtraction is lower than unity.
    \item The flux is at least 3.5\,$\sigma$ above the local sky.
\end{itemize}
We note that: (i) the adopted thresholds are completely arbitrary and chosen empirically; and (ii) some by-hand adjustments to the percentile thresholds were performed to exclude complicated magnitude intervals, e.g., when close to the saturation of the VIS data. Figure~\ref{fig:selections} provides a visual representation of our selections. In the following, we restricted our analysis to only well-measured stars.\looseness=-4

\begin{figure}
    \centering
    \includegraphics[width=\columnwidth]{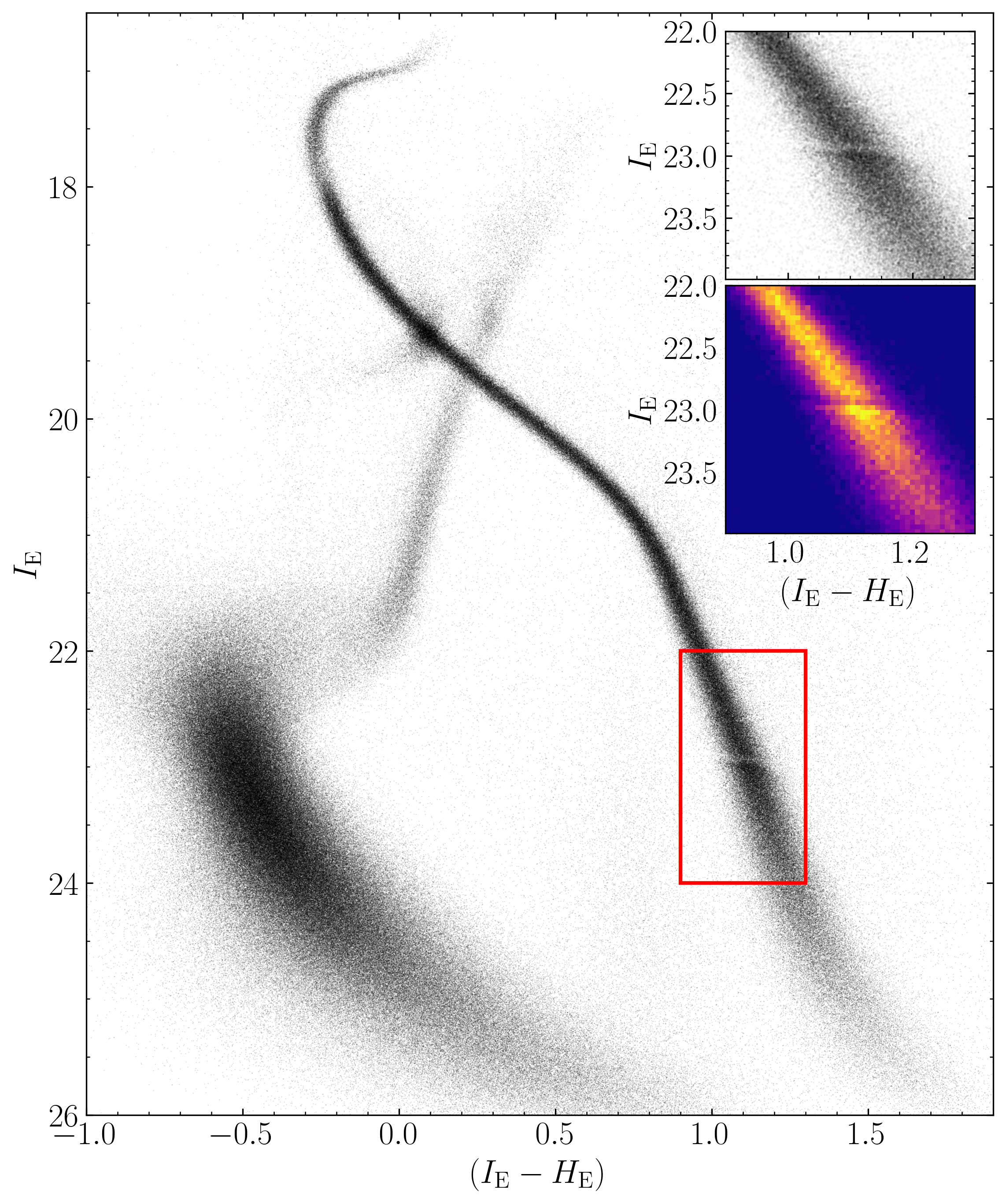}
    \caption{The main panel shows the \IE versus \mbox{$\IE-\HE$} CMD of 47\,Tuc. The narrow sequence corresponds to 47\,Tuc; the broad sequence corresponds to background stars in the SMC. The region highlighted in red is expanded in the top (CMD) and bottom (Hess diagram) insets. The convective gap is clearly visible at \IE\,$\approx$\,22.9.\looseness=-4}
    \label{fig:cmd}
\end{figure}

\begin{figure}
    \centering
    \includegraphics[width=\columnwidth]{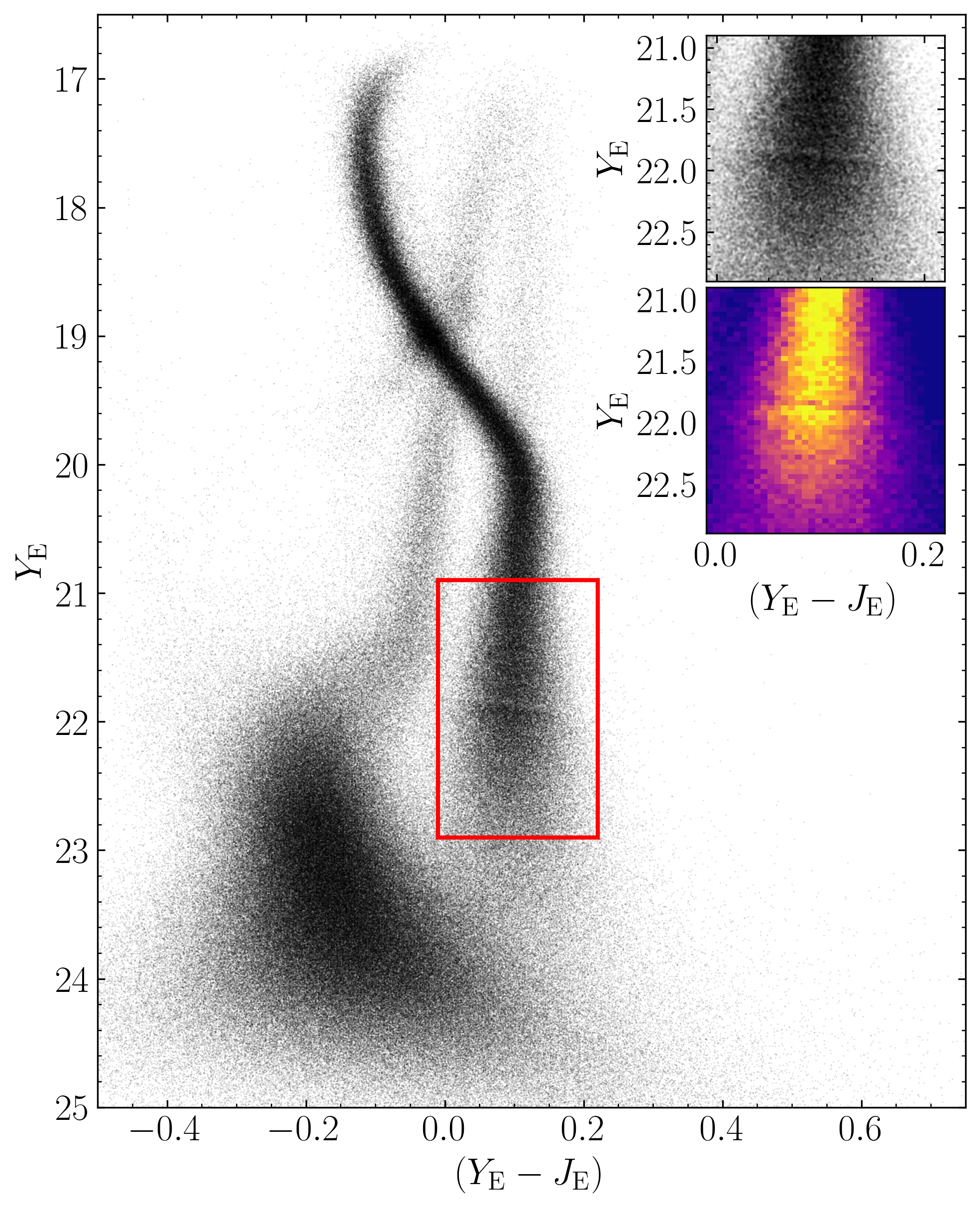}
    \caption{As in Fig.~\ref{fig:cmd} but with the \YE and \JE filters. The convective gap is visible at \YE\,$\approx$\,21.9.\looseness=-4}
    \label{fig:cmdnir}
\end{figure}

Figure~\ref{fig:cmd} shows the \IE versus ($\IE - \HE$) CMD of the stars in this field. The narrow and red sequence that extends from \IE$\approx$\,16.5 to 26 represents 47\,Tuc, whereas stars with a broad and blue distribution are part of the Small Magellanic Cloud (SMC). The red rectangle highlights stars near the convective gap. The zoomed insets clearly show a significant drop in stellar counts around \IE$\approx$\,22.9. Although there are no proper motions to confirm the membership, the stars in this sequence are likely cluster members because SMC stars are bluer than those in 47\,Tuc, and field stars along the same line of sight are expected to be more homogeneously distributed in brightness and position. Furthermore, MS stars within $\pm$0.025 mag of the feature follow the GC's spatial distribution. Finally, the absence of a similar feature in the SMC sequence at the same magnitude level supports the interpretation of an intrinsic gap along the MS of 47\,Tuc, rather than an instrumental error. We applied similar selections to the other NISP filters, and the gap is also visible in purely infrared CMDs (Fig.~\ref{fig:cmdnir}), where the effect of extinction is low, which rules out foreground extinction as a possible cause of the gap.\looseness=-4

We also performed artificial-star tests to check for systematic errors, assess the completeness of our data, and verify the gap we observe along the MS of 47\,Tuc is real. We refer to Appendix~\ref{appendix:artstar} for a detailed description of this part.\looseness=-4

\begin{figure}
    \centering
    \includegraphics[width=\columnwidth]{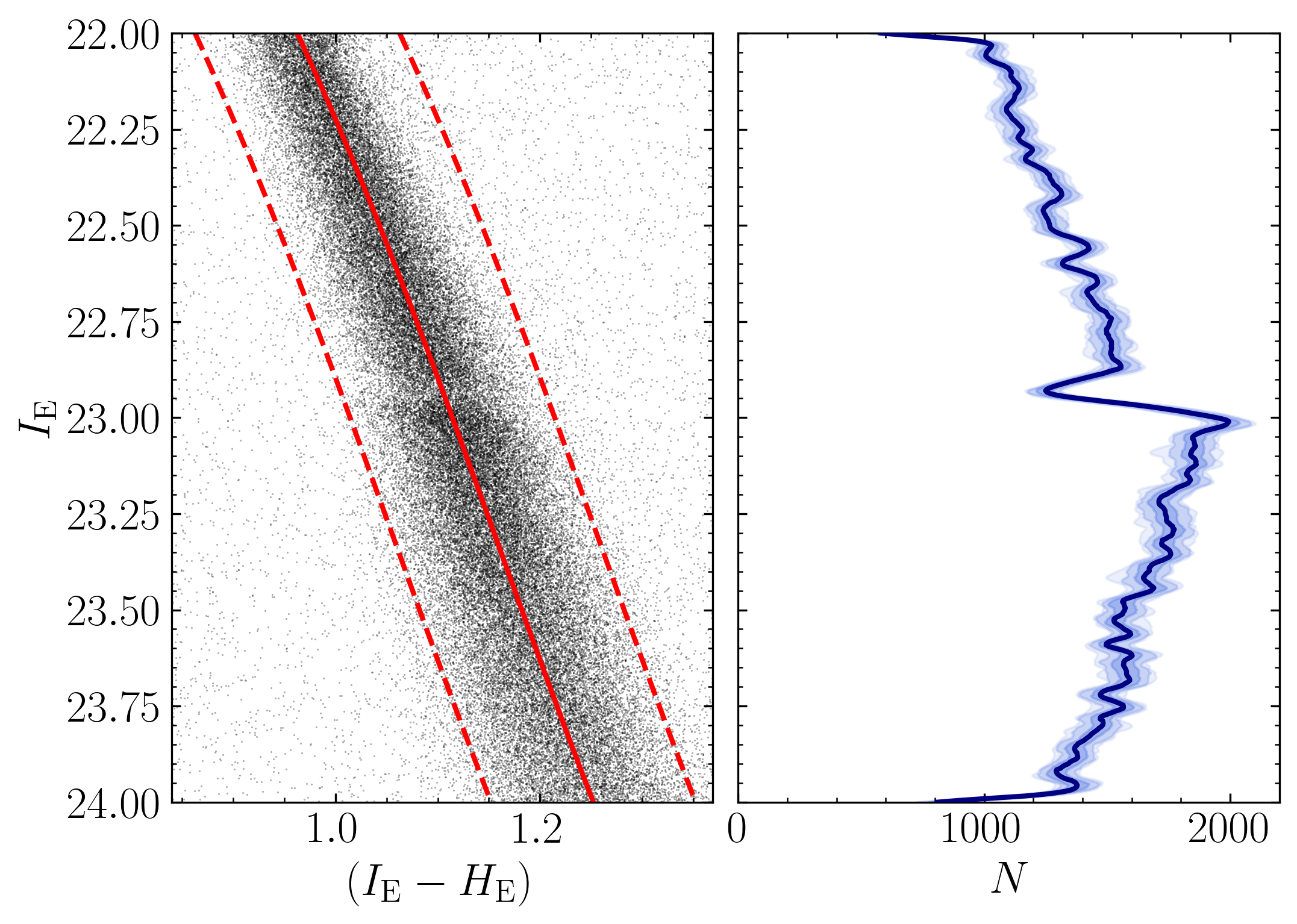}
    \caption{\textit{Left:} \IE versus ($\IE-\HE$) CMD, overlaid with a fiducial line (solid red line), bounded at $\pm$0.1 mag on either side (dashed red lines). \textit{Right:} present\,day LF of 47\,Tuc within these bounds (dark-blue line), with 1$\sigma$, 2$\sigma$, and 3$\sigma$ confidence intervals displayed in lighter shades.\looseness=-4}
    \label{fig:hist}
\end{figure}

\section{Empirical feature validation}\label{sec:validation}

The large sample size makes it possible to validate and quantitatively analyse the convective gap of 47\,Tuc. Following an approach similar to that of \citet{2018ApJ...864..147C}, we initially computed a MS fiducial line as follows. We divided the MS within the red rectangle in Fig.~\ref{fig:cmd} into 30 equally-spaced magnitude bins. In each bin, we used a kernel-density-estimation (KDE) to compute the probability density function (PDF) of the underlying colour distribution, and found the peak of the PDF (the corresponding magnitude was set at the centre of each magnitude bin). The PDF peaks were then smoothed with a spline (the solid line in the left panel of Fig.~\ref{fig:hist}). We then selected all stars within $\pm$0.1 mag in colour of this fiducial line (dashed lines). This threshold was arbitrarily chosen as a compromise between including MS members and excluding binaries and field interlopers. According to our artificial-star tests, the photometric errors for sources in Fig.~\ref{fig:hist} are of $\approx$\,0.017 mag and $\approx$\,0.033 mag for \IE and \HE, respectively.\looseness=-4

We derived the completeness-corrected present-day luminosity function (LF) using an approach that combines averaged-shifted histograms and bootstrap with resampling \citep[e.g.,][]{2019ApJ...873..109L}. The averaged-shifted-histogram technique (i.e., averaging multiple histograms with shifted starting points) removes any bias from the arbitrary choice of the starting point of the histogram; the bootstrap with resampling mitigates potential biases related to the bin size. The averaged-shifted histograms were obtained by averaging ten histograms, shifted by one-tenth of the chosen bin size (i.e., 0.025 mag), and we performed 100 bootstrap realisations. Finally, since the completeness level varies significantly across the field of view, we ran this procedure in six equally populated (12\,400 stars) radial intervals between 8 and 76 arcmin from the centre of 47\,Tuc (we excluded the centremost region because of the very low completeness, see Appendix~\ref{appendix:artstar}), and applied completeness corrections using the outputs of the artificial-star tests for the corresponding radial intervals. Histograms of each radial bin were summed together. The resulting LF is shown in the right panel of Fig.~\ref{fig:hist}. This process allows us to compute the median stellar counts (dark-blue line) and the associated confidence intervals (shaded regions), as well as the magnitude (and its uncertainty) of the convective gap defined from the local minimum of a spline fitted to the LF in the range $22.75 < \IE < 23.25$. We found the local minimum at $\IE^{\rm LF,gap} = 22.921^{+0.067}_{-0.014}$. The uncertainties reported here only include the formal internal errors. The drop in counts at the gap level is not just a statistical fluctuation; the local minimum in the LF at $\IE^{\rm LF,gap}$ remains distinct even at the $3\sigma$ confidence level.\looseness=-4

\section{Comparison with known convective gaps}\label{sec:comparison}

We compared the convective gap in 47\,Tuc with that found in the metal-poor GC NGC\,6397 by \citetalias{griggio2026}. To ensure a meaningful comparison, the photometry of NGC\,6397 was adjusted to account for the differences in distance modulus and reddening relative to 47\,Tuc. For 47\,Tuc, we assumed a distance modulus of 13.24 \citep{2018ApJ...867..132C} and an optical reddening of $E(B-V) = 0.04$ \citep{roman_47tuc}, while for NGC\,6397 we used the values provided in \citetalias{griggio2026}. The $A_\lambda/A_V$ coefficients used in the conversion to absolute magnitudes were taken from the `PAdova TRieste Stellar Evolution Code' \citep[PARSEC;][]{Bressan2012} website.\footnote{\href{https://stev.oapd.inaf.it/cgi-bin/cmd}{https://stev.oapd.inaf.it/cgi-bin/cmd}} These extinction coefficients are derived for a G2V star, using \citet{1989ApJ...345..245C} and \citet{1994ApJ...422..158O} extinction curves with $R_V = 3.1$, which is expected to suffice, given the low $E(B-V)$. To mitigate any uncertainty in the \euclid photometric calibration, we made use of isochrones from the `a Bag of Stellar Tracks and Isochrones' \citep[BaSTI-IAC,][]{Hidalgo2018,Pietrinferni2021} database to offset (in magnitude and colour) the \euclid data until we matched the observed and theoretical MS turn-offs. We assumed an age of 11.5\,Gyr and $[\mathrm{Fe/H}] = -0.75$ from \citet{roman_47tuc} for 47\,Tuc (see also Sect.~\ref{sec:gap_chemical_sensitivity}), while for NGC\,6397 we again referred to the values in \citetalias{griggio2026}. For 47\,Tuc, we found offsets of 0.183 and 0.287 mag in \IE and \HE filters, respectively, whereas for NGC\,6397 the offsets ranged between 0.045 (\IE filter) and 0.233 (\HE filter) mag. The difference in the offsets of the two GCs might be related to the different calibration adopted. As described in Sect.~\ref{sec:datared}, we used the official Level-2 stack calibrated catalogues to calibrate the photometry of 47\,Tuc, whereas \citetalias{griggio2026} calibrated the photometry of NGC\,6397 on the official ERO catalogues \citep{EROGalGCs}. Finally, for NGC\,6397 we selected only the best stars, as described in \citetalias{griggio2026}.

\begin{figure}
    \centering
    \includegraphics[width=\columnwidth]{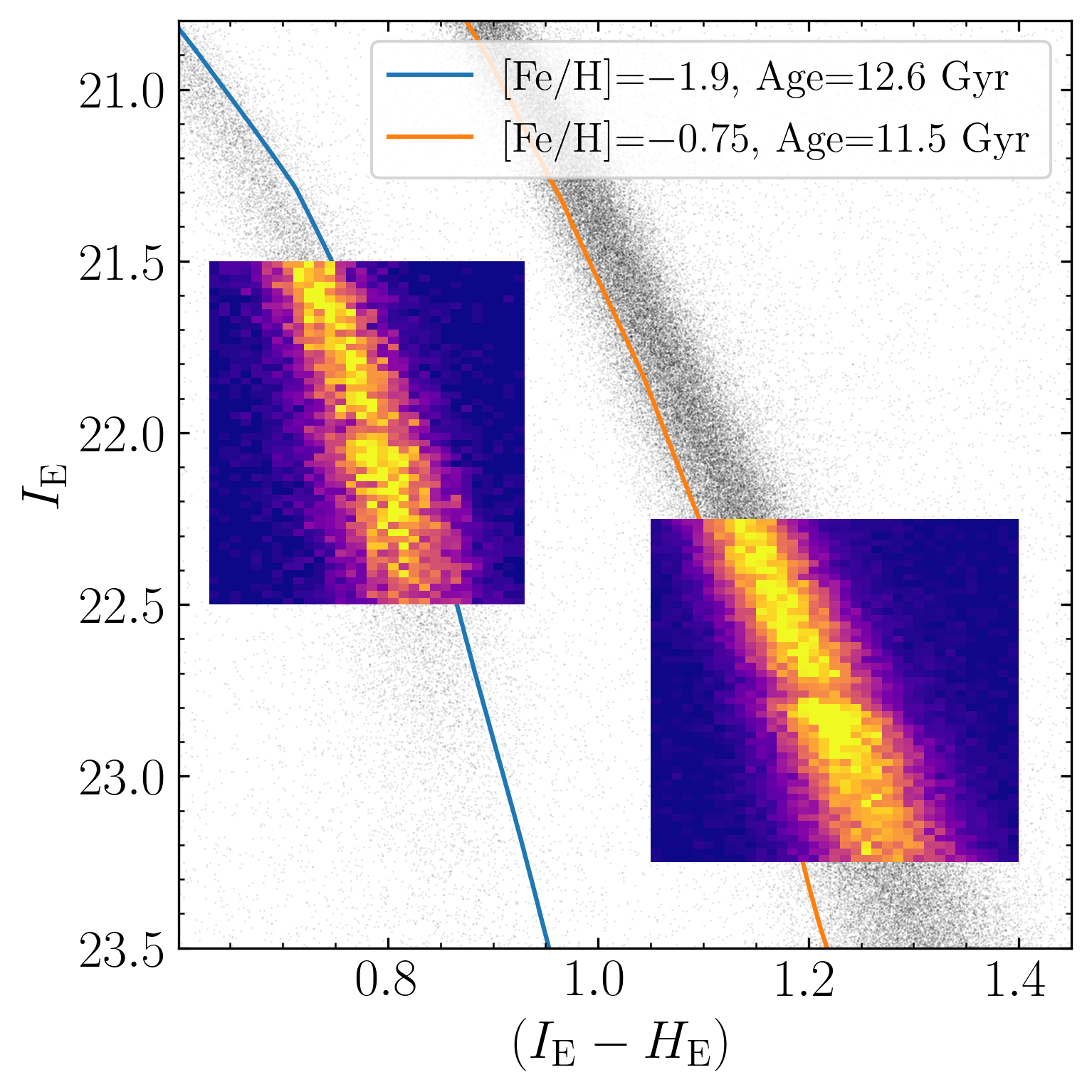}
    \caption{CMD of 47\,Tuc (redder sequence) and NGC~6397 (bluer sequence) in \euclid filters. The photometry in this plot has been rescaled as described in Sect.~\ref{sec:comparison} to ensure a fair comparison between the two systems. For each GC, we plot an Hess diagram to highlight the location and morphology of the convective gap. Two different isochrones (solid lines) are shown for reference.\looseness=-4}
    \label{fig:gcgaps}
\end{figure}

The result is shown in Fig.~\ref{fig:gcgaps}. The Hess diagrams are made with bin sizes and colour scales tailored for each case to best highlight the corresponding gaps. Note that isochrones in Fig.~\ref{fig:gcgaps} typically do not perfectly match the colours of low-mass MS stars. This could be related to shortcomings in the model atmospheres or isochrones in this mass regime (e.g., suboptimal bolometric corrections). Figure~\ref{fig:gcgaps} highlights two primary differences between the convective gaps in these GCs: (i) the gap in the metal-poor cluster NGC\,6397 is $\approx$\,0.75 mag brighter than the gap in the metal-rich 47\,Tuc; (ii) the gap in NGC\,6397 exhibits a steeper slope.\looseness=-4

\begin{figure}
    \centering
    \includegraphics[width=\columnwidth]{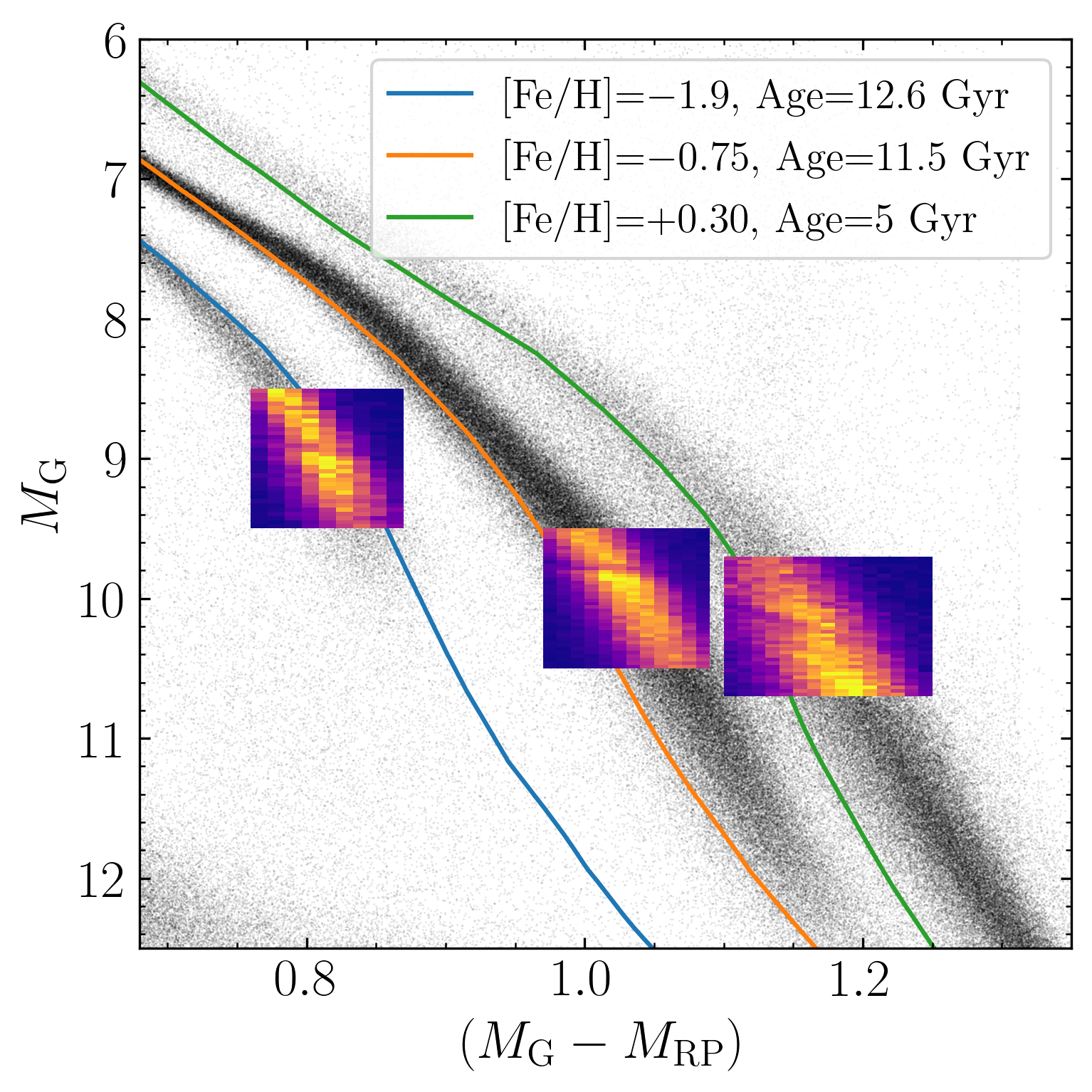}
    \caption{Absolute-magnitude CMD in \gaia filters. The three sequences from the blue to red side of the plot represent NGC\,6397, 47\,Tuc and the \gaia sample, respectively. As in Fig.~\ref{fig:gcgaps}, we also plot an Hess diagram to highlight the gap's location. Solid lines represent three different isochrones for reference.\looseness=-4}
    \label{fig:allgaps}
\end{figure}

We also compared the gaps in these two GCs with the solar neighbourhood gap identified by \citet{2018ApJ...861L..11J}, updating their analysis with the latest \gaia DR3 catalogue. As we lack \euclid photometry for the latter group, we converted our \euclid-based photometry of the two GCs in the \gaia photometric system using BaSTI-IAC isochrones to find appropriate (eighth-order) polynomial colour transformations from \IE and \HE to $G$ and $G_{\rm RP}$ magnitudes (in addition to the aforementioned offsets to compensate for the uncertain photometric calibration). As the \gaia sample is heterogenous in age and metallicity, transforming the \gaia photometry in the \euclid photometric system would have been a more complex task. For the \gaia sample, we used the parallax in the \gaia catalogue and neglected the extinction. We also restricted the analysis to sources within 100\,pc from the Sun that passed the quality selections described in \citet{2022MNRAS.511.4702G} and \citet{2023MNRAS.522L..61S}. The CMD in absolute magnitudes, including the comparison between the three gaps, is provided in Fig.~\ref{fig:allgaps}. Again, position and the slope of the three gaps are different. 

In \gaia absolute magnitudes, the gap of NGC\,6397 is centred at $M_G \approx 9$, and that of 47\,Tuc is again $\approx$\,0.8 mag fainter, similarly to what we reported in the previous \euclid-based comparison between the two GCs. For the \gaia sample, the gap's magnitude ranges from $M_G \approx 10$ (on the blue side of the sequence) to $\approx$\,10.3 (on the red side), as described in \citet{2018ApJ...861L..11J}. Interestingly, the gap's magnitude does not appear to scale linearly with metallicity. The offset between NGC\,6397 and 47\,Tuc ($\Delta [\mathrm{Fe/H}] \approx 1.1$ dex) is $\approx$\,0.8 mag. In contrast, the (super-)solar metallicity \gaia sample -- which is as much more metal-rich than 47\,Tuc as 47\,Tuc itself is more metal-rich than NGC\,6397 -- has a gap which is only 0.2 to 0.5\,mag fainter, and a slope that is once again steeper than 47\,Tuc's. These discrepancies suggest that metallicity alone is not the only driver of the gap's morphology, and that other factors must contribute to morphology of this feature.\looseness=-4

\section{Convective-gap modelling and properties}\label{sec:theory}

The detailed modelling of the complex morphology of the gap and more detailed quantitative conclusions are left to a future, dedicated work. Below, we provide a preliminary analysis of the properties of the gap by comparing the observed LF with theoretical models.

\subsection{Theoretical modelling of the luminosity function}\label{sec:modelling}

Our theoretical LF model is based on the grid of model atmospheres computed in \citet{roman_47tuc} specifically for the parameters of 47\,Tuc, which were partly adopted from the literature and partly determined by the authors fitting their models to the \hst photometry of \citet{2012AJ....143...11K}. In particular, we employed the `ridgeline' model grid from \citet{roman_47tuc}, tailored to the modal oxygen abundance of the cluster, which the authors estimated as $[\mathrm{O/Fe}]=0.33$. All available models were computed for the metallicity of $[\mathrm{Fe/H}]=-0.75$ and a helium mass fraction of $Y=0.25$.\looseness=-4

We computed synthetic photometry in the \euclid \IE band using the \texttt{synphot} utility of \texttt{BasicATLAS} \citep{BasicATLAS}, a distance modulus of 13.24 \citep{2018ApJ...867..132C}, an optical reddening of $E(B-V) = 0.04$ \citep{roman_47tuc}, a total-to-selective extinction of $R_V=3.1$, and the reddening law from \citet{G24_law_0} and \citet{G24_law_software}, but see also \citet{G24_law_1}, \citet{G24_law_2}, \citet{G24_law_3}, and \citet{G24_law_4}.\looseness=-4

We calculated new evolutionary models using Modules for Experiments in Stellar Astrophysics (\texttt{MESA}) code, version \texttt{23.05.1} \citep{MESA,MESA_2,MESA_3,MESA_4,MESA_5}. Our \texttt{MESA} setup was largely adopted from \citet{2024RomanBDs6397}; however, we used the aforementioned model atmospheres from \citet{roman_47tuc} as the outer boundary conditions, as well as a different mass-sampling algorithm. All evolutionary calculations were carried out to the target age of $11.5\ \mathrm{Gyr}$.

The stellar mass range near the convective gap exhibits three distinct regimes of stellar evolution, described in detail by \citet{convective_kissing}. At masses below some threshold value $M_{\rm C}$, the star remains fully convective at the target age. At masses above another threshold value $M_{\rm R}>M_{\rm C}$, the star arrives at the target age with a stable radiative zone in its interior that separates the convective core from the envelope. The third regime is found at stellar masses between $M_{\rm C}$ and $M_{\rm R}$, and represents a transition phase between those two stable states. In this transition region, the star oscillates between sustaining a radiative zone and brief episodes of full convection that are known as the `convective kissing instability'.

To choose appropriate initial stellar masses for our \texttt{MESA} models, we first determined the values of $M_{\rm C}$ and $M_{\rm R}$ by coupling \texttt{MESA} with a bisection root-finding method. We calculated $M_{\rm C}$ and $M_{\rm R}$ as $0.3317\ M_\odot$ and $0.3358\ M_\odot$, respectively. We then computed our final grid of evolutionary models using the following mass-sampling algorithm. For masses $M$ in the ranges $0.3\ M_\odot\leq M\leq M_{\rm C}$, and $M_{\rm R}\leq M\leq 0.4\ M_\odot$, we used the adaptive mass-sampling algorithm from \citet{roman_47tuc}, which keeps luminosity differences between adjacent models under $0.005\ \mathrm{dex}$. Within the transition region ($M_{\rm C}\leq M\leq M_{\rm R}$), the mass--luminosity relationship is expected to have abrupt discontinuities near stellar masses that experience convective kissing instabilities near the target age. To capture the full complexity of this behaviour, we used a dense uniform sampling with the mass difference between adjacent models of $2 \times 10^{-5}\ M_\odot$.

In our \texttt{MESA} models, episodes of full convection associated with convective kissing instabilities persist for only a single time step, and are followed by abrupt, discontinuous changes in stellar parameters. We attempted to resolve these discontinuities by reducing the maximum allowed time step; however, they remained present even at very small step sizes. Furthermore, requiring the inter-step variations in stellar parameters to be smaller than those discontinuities resulted in \texttt{MESA} being entirely unable to evolve the star past the instability. These numerical difficulties suggest that, at least in the regime explored here, the oscillatory behaviour of stars in the transition region may be partly driven by modelling artefacts. Physically, one might instead expect stars in this regime to approach a quasi-stable intermediate state between fully convective and partly radiative configurations, potentially mediated by more sophisticated processes such as semi-convection \citep{2013A&A...552A..76S}.

\begin{figure}
    \centering
    \includegraphics[width=\columnwidth]{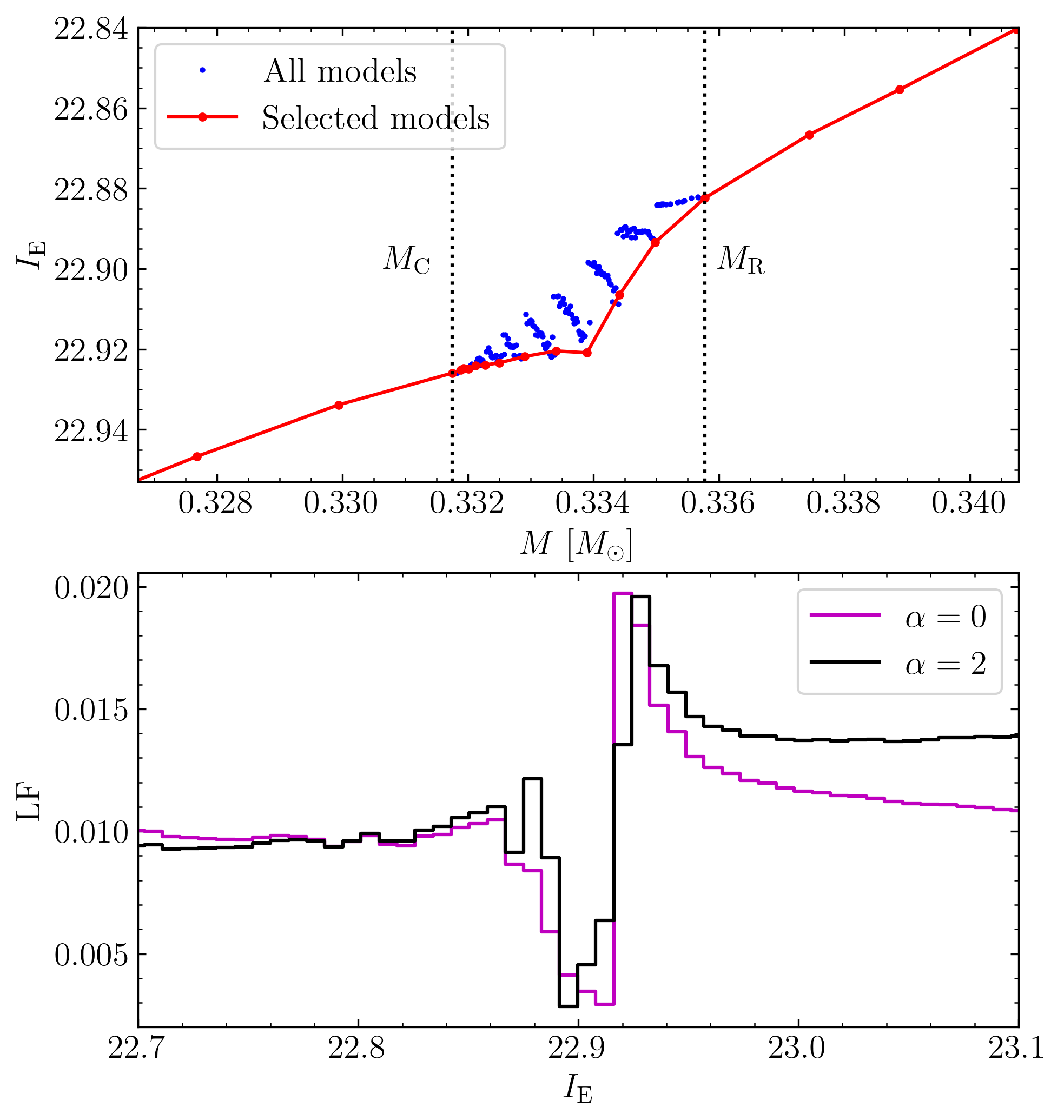}
    \caption{{\it Top:} theoretical mass-luminosity relationship near the convective gap of 47\,Tuc. The blue markers represent all \texttt{MESA} models in the calculated grid, while the final mass-luminosity relationship was constructed from a subset of those models (red line) that are either outside the transition region or have experienced a convective kissing instability just before reaching the target age. The transition region is bound by threshold masses $M_{\rm C}$ and $M_{\rm R}$, as described in the text. {\it Bottom:} two examples of normalised theoretical LFs (see the text for details) computed in this study near the convective gap for different values of the MF slope $\alpha$.\looseness=-4}
    \label{fig:lfmodel}
\end{figure}

While developing a fully self-consistent model of this behaviour is beyond the scope of this paper, we adopted a simplified prescription to approximate it. Specifically, we assumed that this intermediate state is best represented by the stellar structure immediately following an episode of full convection. Accordingly, within the transition region we retain only those \texttt{MESA} models that have undergone a convective instability immediately prior to reaching the target age, i.e., models for which the total number of instability episodes would decrease by one at the next mass step. We also explored alternative selections, including retaining all models in the transition region and selecting only those immediately preceding an instability; however, both approaches lead to increased scatter and visible discontinuities in the mass-luminosity relationship, indicating a stronger sensitivity to mass sampling. The adopted selection has the key advantage of producing a mass–luminosity relationship that is minimally affected by such artefacts. Our final mass-luminosity relationship is shown in the upper panel of Fig.~\ref{fig:lfmodel}.\looseness=-4

The theoretical mass–luminosity relationship was converted into model LFs using a Monte Carlo approach. In each of $1000$ realisations, we drew $10^6$ stellar masses from a power-law mass function, parametrised by the slope $\alpha$ as $P(M)\propto M^{-\alpha}$. These masses were then mapped to \IE magnitudes via interpolation of the theoretical mass–luminosity relationship and subsequently binned into $100$ uniform magnitude bins. The final LF was obtained by averaging the bin counts across all realisations. Examples of model LFs for $\alpha=0$ and $\alpha=2$ are provided in the lower panel of Fig.~\ref{fig:lfmodel}. We found the gap's centroid at about 0.33 $M_\odot$ and $\IE = 22.903$, in agreement with the empirical value from Sect.~\ref{sec:validation}, within the uncertainties.\looseness=-4

\subsection{Physical properties}\label{sec:gap_chemical_sensitivity}

The gap's magnitude is expected to depend on chemical composition through changes in both stellar structure (which shift the mass at which the gap occurs) and the spectral energy distribution of the atmosphere (which alters the mapping between the mass of the gap and its magnitude). We estimate the sensitivity of the convective-gap magnitude to the bulk metallicity of the cluster $[\mathrm{Fe/H}]$, as well as oxygen-to-iron abundance $[\mathrm{O/Fe}]$, using isochrones from the literature. Isolating oxygen is especially important, since it is both the dominant metal and exhibits large star-to-star variations in most GCs \citep{2018ARA&A..56...83B}. We used isochrones from the BaSTI-IAC database to probe the relationship between the magnitude of the gap and metallicity, and the isochrones from \citet[][henceforth referred to as G24 isochrones]{roman_47tuc} to probe the equivalent relationship with the oxygen abundance.\looseness=-4

While BaSTI-IAC isochrones do not sample stellar mass finely enough to resolve the convective gap directly, they do capture the associated discontinuity in the slope of the mass–luminosity relation \citep{1997MNRAS.287..402K}. We therefore define a characteristic magnitude of this feature as the point where $\left| \diff^2 \IE/{\diff}M^2 \right|$ is maximised over the interval $0.3 \leq M/M_\odot \leq 0.4$, and assume that this magnitude is correlated with the location of the gap. Comparing BaSTI-IAC isochrones with $[\mathrm{Fe/H}]=-0.75$ and $-0.70$, we find that the slope discontinuity (and, by extension, the magnitude of the gap) becomes fainter by 0.031\,mag per 0.1\,dex increase in $[\mathrm{Fe/H}]$.

The G24 isochrones are even more limited in mass resolution, preventing a direct identification of the slope discontinuity. To address this, we supplemented the G24 isochrones at $[\mathrm{O/Fe}]=0.33$ and $0.48$ with new evolutionary models and synthetic photometry in \Euclid filters, computed following the procedure described in Sect.~\ref{sec:modelling}. We then identified the mass bounds of the convective transition, and approximated the mass of the gap as their midpoint. Interpolating the corresponding G24 isochrones to the estimated mass of the gap then yielded the associated gap's magnitudes. We found that the gap becomes fainter by 0.026 mag per 0.1\,dex increase in the $[\mathrm{O/Fe}]$.\looseness=-4

We also found that the mass of the gap itself is only weakly sensitive to chemical composition, varying by $<$\,0.001\,$M_\odot$ per 0.1\,dex change in either $[\mathrm{Fe/H}]$ or $[\mathrm{O/Fe}]$. This suggests that most of the chemical sensitivity of the gap is due to the mass-luminosity relationship, not variations in the stellar mass corresponding to the gap.\looseness=-4

The similarity of these two magnitudes suggests that oxygen may act as the dominant determinant of the \IE gap's magnitude in 47\,Tuc. This contrasts with results for the more metal-poor GC NGC\,6397 (\citetalias{griggio2026}), where the influence of oxygen is significantly weaker than that of the overall metallicity. This result suggests that, in 47\,Tuc, the primary driver of the chemical sensitivity of the gap may be the spectral energy distribution, which is strongly affected by light-element abundances \citep[e.g.,][]{2012ApJ...755...15V}. Such abundance variations are a defining characteristic of multiple stellar populations (mPOPs) in GCs \citep{2020CassimPOPsRv}. The pronounced sensitivity of the gap to light-element abundances suggests that it may provide a novel and independent probe of this phenomenon.\looseness=-4

Given the dependence of the location of the gap on chemical abundances, the mono-metallic gap profile in Fig.~\ref{fig:lfmodel} will be shifted in magnitude if the mean chemistry of 47\,Tuc differs from the adopted model value. If the resulting morphology is understood as a superposition of gaps with slightly offset magnitudes or, equivalently, as the convolution of the mono-metallic gap profile with the chemical dispersion mapped into magnitude space, then the magnitude and broadening of the gap will encode information about the star-to-star variations in chemistry of 47\,Tuc (noting that the offset is also sensitive to photometric zero-points, distance, and reddening). In principle, this provides a sensitive probe of mPOPs, potentially offering higher precision than spectroscopic measurements.\looseness=-4

\begin{figure}
    \centering
    \includegraphics[width=\columnwidth]{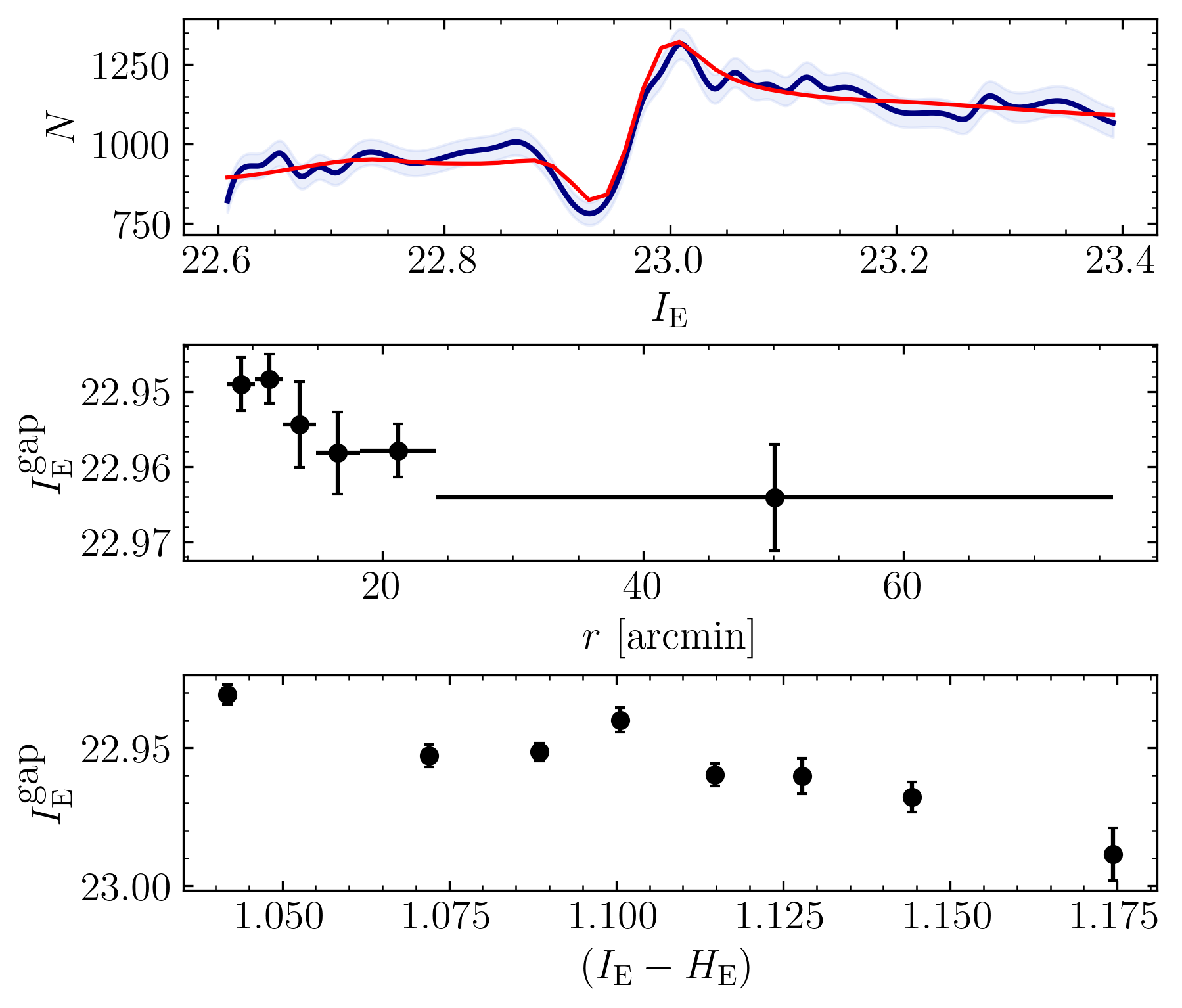}
    \caption{{\it Top:} smoothed present-day LF of 47\,Tuc (blue line) fitted with a mono-metallic theoretical gap model, convolved with a Gaussian kernel to account for an $[\mathrm{Fe/H}]$ spread (red line). The model was fitted without binning. {\it Middle:} radial variation of $\IE^{\rm gap}$. Horizontal lines indicate the coverage of each bin. {\it Bottom:} colour-dependent ($\IE-\HE$) variations of $\IE^{\rm gap}$ across the width of the MS.\looseness=-4}
    \label{fig:theory}
\end{figure}

As a proof of concept, we fitted the theoretical gap profile to the observed LF in the range $22.6 < \IE < 23.4$. We allowed the fitter to apply arbitrary magnitude offsets and Gaussian broadening to the gap profile. The latter was parametrised by the standard deviation of the Gaussian kernel. Overall, we constrained three free parameters: the mass-function (MF) slope ($\alpha$), where $P(M)\propto M^{-\alpha}$, the central gap's magnitude $\IE^{\rm gap}$, and the broadening width $\sigma$. The model fit was carried out by treating the profile of the gap as a probability distribution function and maximising the likelihood of the observed magnitudes using the Goodman--Weare \citep{MCMC} Markov chain Monte Carlo (MCMC) algorithm, implemented in the \texttt{emcee} package \citep{2013PASP..125..306F}. Our model fit is shown in the top panel of Fig.~\ref{fig:theory}. The best-fit parameters are estimated as $\IE^{\rm gap} = 22.954 \pm 0.002$, $\sigma = 0.025 \pm 0.001$\,mag, and $\alpha = 0.61 \pm 0.05$. Because $\sigma$ is a relative measure, its best-fit value is insensitive to systematic photometric offsets. Note that these uncertainties are purely statistical; they do not include systematic errors introduced by stellar modelling, which could make the total uncertainties substantially larger.\looseness=-4

We repeated the analysis in the same individual radial bins defined in Sect.~\ref{sec:validation} (middle panel of Fig.~\ref{fig:theory}). The results tentatively suggest that the convective gap is brighter in the inner regions of the cluster, which would correspond to lower oxygen content. We also computed the magnitude of the gap in eight equally populated colour bins\footnote{We fitted the gap magnitude as a function of colour with a straight line, and found a slope of $0.34 \pm 0.08$. Using a similar but more robust methodology, \citetalias{griggio2026} found a slope of $0.5713 \pm 0.0014$ for NGC~6397. This result further corroborates the discussion in Sect.~\ref{sec:comparison}.} (bottom panel of Fig.~\ref{fig:theory}), and found a possible non-monotonic feature (`wiggle') near $\IE-\HE \approx 1.1$.\looseness=-4

While we refrain from detailed quantitative conclusions and defer systematic error analyses to future studies, we identify several tentative mPOP signatures in 47\,Tuc reflected in its gap.\looseness=-4
\begin{itemize}
    \item We find that the gap is broadened by $\sigma = (0.025 \pm 0.001)$\,mag compared to the mono-metallic gap profile. By subtracting the measurement errors in quadrature (see Sect.~\ref{sec:validation}) from the fitted broadening, we can estimate the intrinsic broadening $\sigma_{\rm int} \sim 0.018$ mag. As argued above, the gap's location and broadening are likely driven by light-element abundances (notably oxygen). We thus attribute this broadening to the presence of mPOPs.\looseness=-4
    \item Light-element abundances also help determine the colour of low-mass stars, hence the relation between the colour and magnitude of the gap (i.e., the slope) in the CMD (bottom panel of Fig.~\ref{fig:theory}) can identify opacity and evolutionary changes caused by differing abundances among mPOPs. Such relationships may be unveiled in a future 2D analysis of the full CMD.\looseness=-4
    \item The tentative `wiggle' in the gap near $\IE-\HE \approx 1.1$, if confirmed, hints at multi-modality in light-element dispersions, as distinct mPOPs would have distinct relationships between elements that affect the magnitude of the gap.\looseness=-4
    \item The magnitude of the gap is expected to shift by about 0.026\,mag per 0.1\,dex increase in $[\mathrm{O/Fe}]$, which is a value remarkably similar to the observed gap's width, $\sigma = (0.025 \pm 0.001)$\,mag. If oxygen variations alone drive the broadening, this implies an $[\mathrm{O/Fe}]$ dispersion of about 0.1\,dex, which is substantially smaller than spectroscopic measurements (e.g., \citealt{nominal_T14}). This discrepancy could suggest \mbox{(anti-)correlations} between $[\mathrm{O/Fe}]$ and other elements (typical of GC mPOPs) influencing the magnitude of the gap, effectively reducing the net photometric spread. For example, \citetalias{griggio2026} showed that the magnitude of the gap shifts to fainter values by about 0.015 mag per 0.01 increase in $Y$, meaning that an increase in either $Y$ or $[\mathrm{O/Fe}]$ should make the gap fainter. As second-population stars are O-poor, the increase in $Y$ could compensate for the effects of light-element abundance variations. Photometric investigations of 47\,Tuc have revealed that this cluster hosts two populations with an average $Y$ difference of $0.011 \pm 0.005$ and a maximum $Y$ variation of $0.049 \pm 0.005$ \citep{2018MNRAS.481.5098M}. Thus, we could speculate that $Y$ has impact in the morphology of the gap. However, since we currently lack detailed theoretical support for this interpretation, we defer to future, dedicated follow-up works any firm conclusion.\looseness=-4
    \item Finally, the middle panel of Fig.~\ref{fig:theory} suggests that the inner regions of 47\,Tuc are more oxygen-poor, which agrees with the radial trends of the mPOPs in 47\,Tuc \citep{2025MNRAS.536.1077M}.\looseness=-4
\end{itemize}
Our preliminary analysis indicates that the convective gap, and its detailed morphology as revealed by \euclid, as a powerful diagnostic of the internal chemical structure of GCs.\looseness=-4

\section{Conclusions}

The convective gap in the lower MS of 47\,Tuc manifests itself as a clear drop in the present-day LF at $\IE = 22.954 \pm 0.002$ (internal errors). This represents the second detection of this feature in a GC -- the clearest to date, enabled by \euclid's field of view, resolution, and precision. The \euclid-based analyses establish the convective gap as a CMD feature identifiable with very high precision. While the underlying physics requires further characterisation, the magnitude of the gap has the potential to become a new, more precise standard candle, provided that standard uncertainties -- such as differential reddening and average metallicity -- are adequately accounted for (\citetalias{griggio2026}).\looseness=-4

We carried out a preliminary analysis of the properties of the gap by comparing the observed LF with theoretical models. The gap and its morphology can be important tools to probe the chemical composition of GCs, in particular metallicity and light-element abundances. Intriguingly, the gap broadening and location seem driven by the abundance of light elements, in particular oxygen, which in 47\,Tuc has a significant spread \citep{nominal_C14,2024ApJ...969L...8M,roman_47tuc,2025A&A...694A..68S}. In addition, the observed radial variation of the convective-gap magnitude $\IE^{\rm gap}$ is remarkably similar (decrease out to about 20 arcmin and then a constant plateau) to the radial trend of the fraction of oxygen-poor stars in 47\,Tuc observed by \citet{2025MNRAS.536.1077M}. These findings suggest that the gap could be a sensitive tool for investigating mPOPs with potentially higher precision than spectroscopic measurements. However, a more detailed modelling is still needed to better understand this feature.\looseness=-4

The convective gap of 47\,Tuc provides an important bridge between the metal-poor gap of NGC\,6397 (\citetalias{griggio2026}), and the \mbox{(super-)solar-metallicity} gap of the stars in the solar neighbourhood \citep{2018ApJ...861L..11J}. The comparisons shown in Figs.~\ref{fig:gcgaps} and \ref{fig:allgaps} seem to support our preliminary interpretation of the gap properties. Indeed, despite uncertainties in photometric transformations, the gap's position and slope vary non-linearly with metallicity, indicating that other factors play a crucial role in shaping it. Future observations with \euclid will be instrumental in mapping this subtle feature across a broader range of stellar populations, further refining our understanding of the internal structure of low-mass stars.\looseness=-4

\section*{Data availability} The \euclid Level-2 data of 47\,Tuc used in our work are available at the CDS via anonymous ftp to \href{cdsarc.u-strasbg.fr}{cdsarc.u-strasbg.fr} (130.79.128.5) or via \href{http://cdsweb.u-strasbg.fr/cgi-bin/qcat?J/A+A/}{http://cdsweb.u-strasbg.fr/cgi-bin/qcat?J/A$+$A/}.

\begin{acknowledgements}
The authors thank the anonymous referee
for the thoughtful suggestions that improved the quality of the
paper.
\AckEC
This work has made use of data from the European Space Agency (ESA) mission {\it Gaia} (\url{https://www.cosmos.esa.int/gaia}), processed by the {\it Gaia} Data Processing and Analysis Consortium (DPAC, \url{https://www.cosmos.esa.int/web/gaia/dpac/consortium}). Funding for the DPAC has been provided by national institutions, in particular the institutions participating in the {\it Gaia} Multilateral Agreement. We made use of \texttt{astropy}, a community-developed core \texttt{python} package for Astronomy.\looseness-4
\end{acknowledgements}

\bibliography{Euclid_47Tuc}

\begin{appendix}

\section{Artificial-star test}\label{appendix:artstar}

\begin{figure}
    \centering
    \includegraphics[width=\columnwidth]{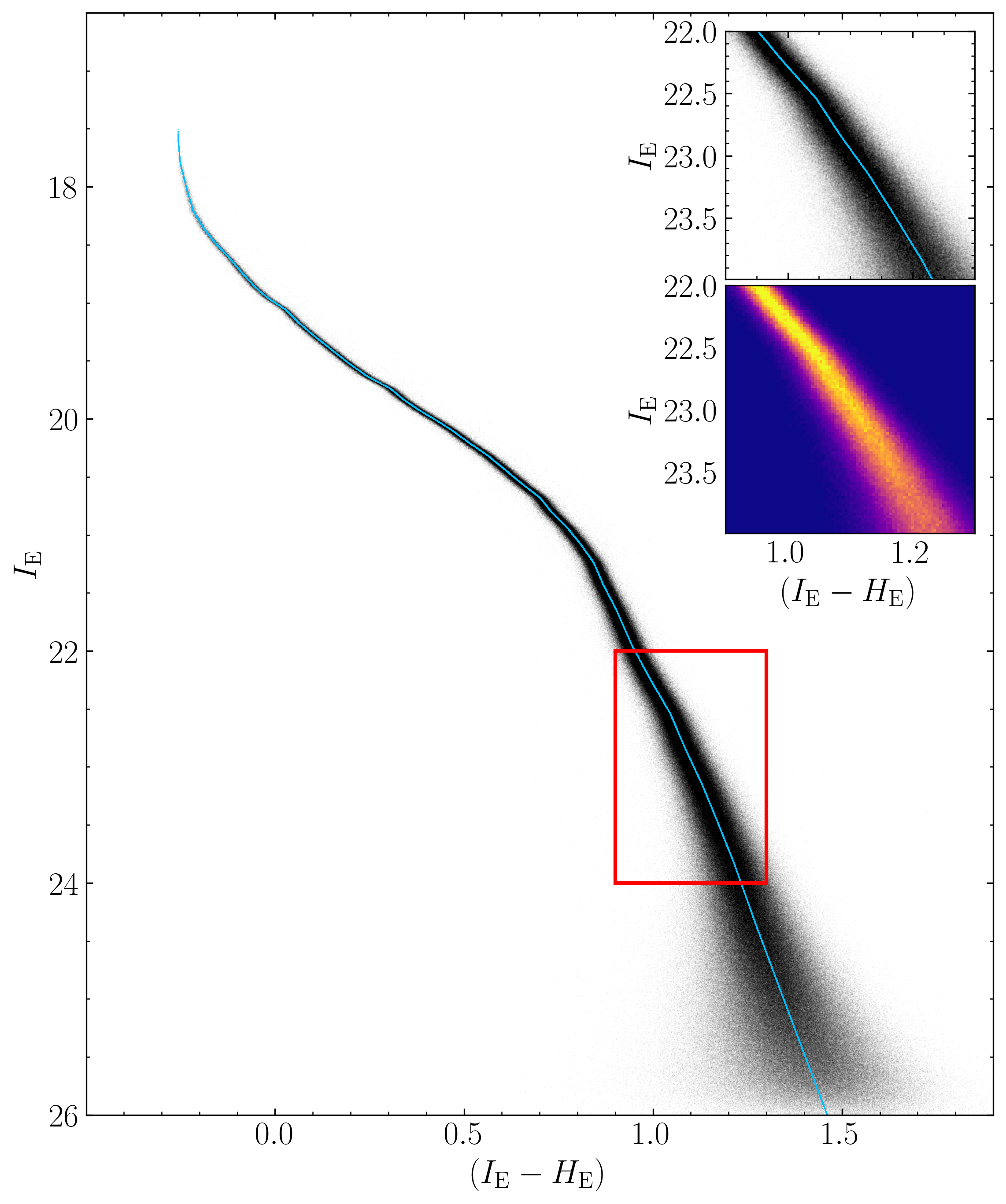}
    \caption{Similar to Fig.~\ref{fig:cmd} but using artificial stars. The light-blue lines represent the input magnitudes and colours for the artificial-star test, whereas black points are recovered stars that passed the same quality selections applied to the real data (see the text for details). No gap is present in these synthetic CMDs.\looseness=-4}
    \label{fig:cmdart}
\end{figure}

We performed artificial-star tests to assess completeness and accuracy. We generated 10$^7$ artificial stars with random magnitudes that follow an LF with an exponential sampling (i.e., the number of stars exponentially increase towards the faint-end of the distribution). We associated random positions selected from either a flat (like that of field objects) or Gaussian distributions (centred in the centre of 47\,Tuc and with $\sigma$ of 5000 pixels) with a 1:1 ratio. We fed this list of artificial stars to \kstwo. \kstwo added each source one at a time to each \euclid image, measured it analogously to real stars, and then removed the object from the images so as not to bias the measurement of the other artificial stars. Figure~\ref{fig:cmdart} shows the result of the artificial-star test. This figure is an analogue of Fig.~\ref{fig:cmd}, but using artificial sources. The absence of any gap in our synthetic CMD confirms that the observed discontinuity the LF of 47\,Tuc is not an artefact.

We compared the input and output magnitudes of the artificial stars to search for the presence of systematic errors \citep[as in, e.g.,][]{2009ApJ...697..965B,2024A&A...690A.371L}. We found no clear trend, in concordance with the analysis of \citetalias{griggio2026} for the \euclid data of NGC\,6397.

\begin{figure}
    \centering
    \includegraphics[width=\columnwidth]{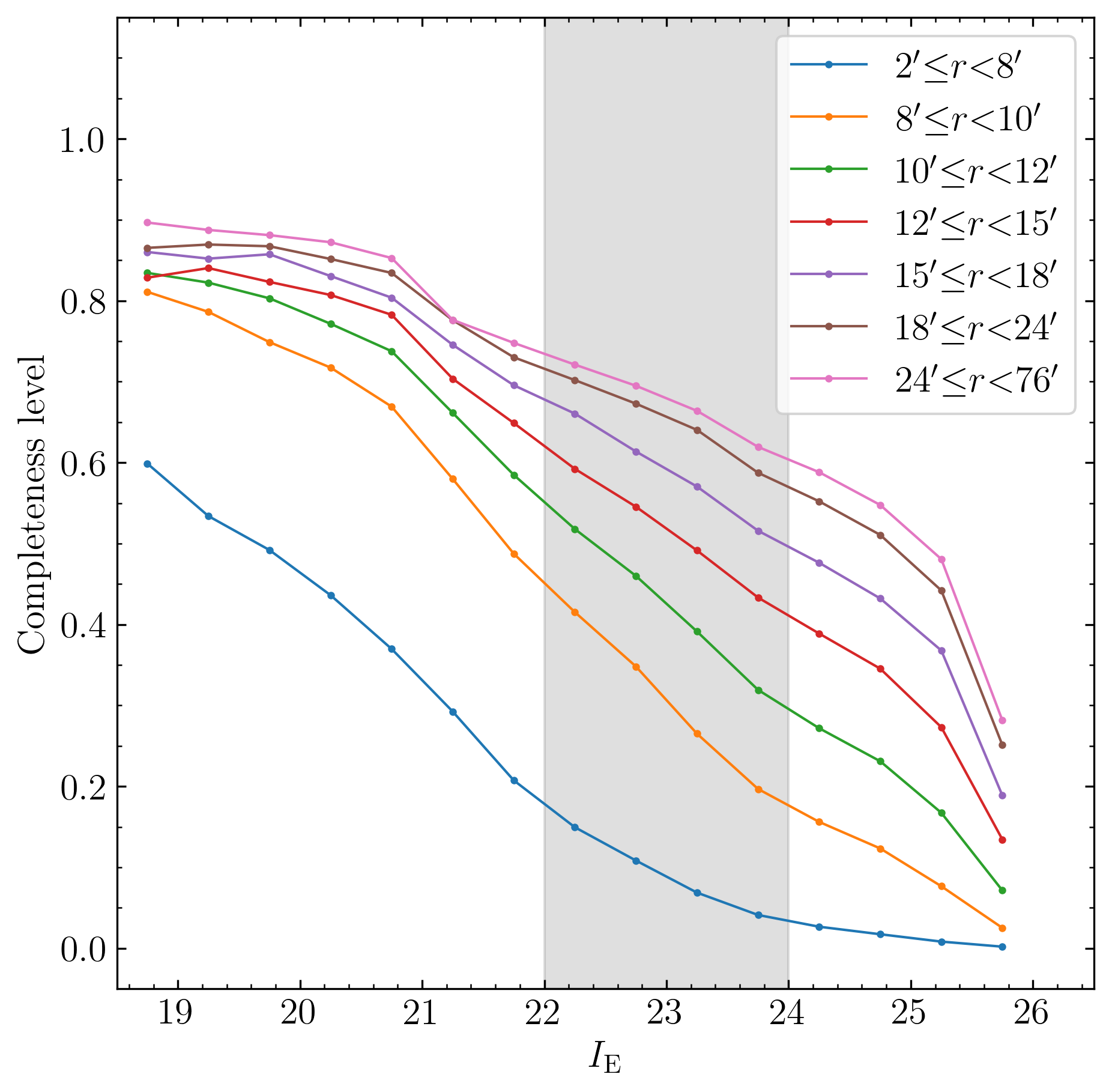}
    \caption{Completeness level as a function of \IE. The grey area highlights the magnitude interval used to compute the present-day LF in the region of the MS that contains the convective gap.}
    \label{fig:completeness}
\end{figure}

We also estimated the completeness of our data. Figure~\ref{fig:completeness} presents an overview of the completeness level in the field of 47\,Tuc that were used to correct the observed LF (see Sect.~\ref{sec:validation}). We considered an artificial star as recovered if the input and output positions agree to within half a pixel, the input and output magnitudes differ by less than 0.75\,mag (i.e., a factor of 2 in flux), and the artificial star passes all photometric criteria described before. We split the field into seven regions, using the same radial boundaries used for real stars in Sect.~\ref{sec:validation}. The level of crowding is important in this data set, and the NISP data of the centremost region suffer the most from incompleteness. For this reason, in the analysis of the LF of 47\,Tuc, we excluded all stars within 8 arcmin of the cluster's centre.\looseness=-4

\end{appendix}


\end{document}